\documentclass[english,structabstract]{aa}
\usepackage{mathptmx}
\usepackage[T1]{fontenc}
\usepackage[latin9]{inputenc}
\usepackage{array}
\usepackage{url}
\usepackage{amstext}
\usepackage{amssymb}
\usepackage{graphicx}
\usepackage[authoryear]{natbib}

\makeatletter

\providecommand{\tabularnewline}{\\}

\usepackage{txfonts}

\makeatother

\usepackage{babel}
\begin{document}

\title{Surface chemistry in the interstellar medium }

\subtitle{I - $\mathrm{H}_{2}$ formation by Langmuir-Hinshelwood and Eley-Rideal
mechanisms}

\author{Jacques Le Bourlot\inst{1}\and Franck Le Petit\inst{1}\and Cecilia
Pinto\inst{1}\and Evelyne Roueff\inst{1}\and Fabrice Roy\inst{1}}

\institute{LUTH, Observatoire de Paris, CNRS, Universit\'e Paris Diderot ; 5 place
Jules Janssen 92190 Meudon, France\\
\email{jacques.lebourlot@obspm.fr}}

\abstract{$\mathrm{H}_{2}$ formation remains a major issue for the understanding
of interstellar physics} {We want to include as much experimental
and theoretical information as possible to describe $\mathrm{H}_{2}$
formation in astrophysical environments.} {We have modified the Meudon
PDR code to include detailed treatment of $\mathrm{H_{2}}$ formation
mechanisms: i) Langmuir-Hinshelwood mechanism taking into account
the contribution of the different sizes of dust grains in the diffusion
processes and ii) the Eley-Rideal mechanism.} {We are able in this
way to form $\mathrm{H}_{2}$ even in regions where dust temperature
is larger than $25\,\mathrm{K}$. We also show that formation by Eley-Rideal
mechanism can be a significant source of heating of the gas. We derive
line intensities for various astrophysical conditions.} {Such an
approach results in an enhanced $\mathrm{H_{2}}$ formation rate compared
to the ``standard'' $3\times10^{-17}\, n_{\mathrm{H}}\, n(\mathrm{H)\,}\mathrm{cm}^{3}\,\mathrm{s}^{-1}$
expression\textbf{.}}

\keywords{astrochemistry - ISM: general - ISM: clouds - ISM: molecules}

\date{\label{dat:Received-...;-Accepted...}}

\maketitle

\section{Introduction\label{sec:Introduction}}

Despite the general agreement that $\mathrm{H}_{2}$ formation occurs
on grain surfaces, the actual formation mechanisms at work in interstellar
conditions are still under debate. We present in this paper a new
chemical model of\textbf{ }$\mathrm{H}_{2}$ formation\textbf{ }which
encompasses recent laboratory experimental and theoretical results
on this topic\textbf{ }and show some consequences on model results\textbf{.
}A follow up paper will extend these results to other molecules on
grains\textbf{.} We recall in Sect.~\ref{sec:Historical-status}
the previous assumptions concerning $\mathrm{H}_{2}$ formation, in
relation with observational issues. Sect.~\ref{sec:Model-calculations}
describes two new implementations concerning i) the contribution of
different dust grain sizes to the formation process within the Langmuir-Hinshelwood
mechanism (labelled LH in the following), ii) the introduction of
the Eley-Rideal mechanism (labelled as ER) for $\mathrm{H}_{2}$ formation.
We display and compare the results obtained for different pressures
and radiation fields in Sect.~\ref{sec:Model-calculations} pertaining
both to diffuse gas and dense Photon Dominated Regions (PDRs) conditions.
Particular focus is given on the values of the $\mathrm{H}_{2}$ formation
rates themselves and on observational tests such as the emissivities
of fine structure transitions of the abundant atoms and ions and infrared
$\mathrm{H}_{2}$ transitions. We present our conclusions in Sect.~\ref{sec:Conclusions}.

\section{Historical status\label{sec:Historical-status}}

\subsection{$\mathrm{H}_{2}$ formation efficiency\label{sub:Old_form}}

\citet{hollenbach:71}\textbf{ }have estimated the binding energies
of $\mathrm{H}$ and $\mathrm{H_{2}}$ on various surfaces as well
as the corresponding sticking efficiencies and have subsequently developed
the first theory of the formation mechanism of $\mathrm{H}_{2}$ on
dust grain surfaces in interstellar conditions due to $\mathrm{H}$
atoms diffusion at the surface of the grains.\textbf{ }The formation
rate has subsequently been derived thanks to Copernicus VUV observations
of $\mathrm{H}_{2}$ in absorption in diffuse interstellar clouds
\citep{jura:74}. This mean formation rate, confirmed by FUSE observations
(\citet{gry:02}), is $3\times10^{-17}\, n_{\mathrm{H}}\, n(\mathrm{H})\,\mathrm{cm}^{3}\,\mathrm{s}^{-1}$.
The launch of ISO and the development of infrared astronomy on ground
based telescopes has allowed detections of warm excited $\mathrm{H}_{2}$
via its electric quadrupolar transitions in the infrared in Photon
Dominated Regions (PDRs). At long wavelengths ($\sim\,7-28\,\mu\mathrm{m}$)
pure rotational transitions are taking place and are detected thanks
to spatial instruments whereas rovibrational transitions at shorter
wavelengths can be observed from the ground in the I, J, H and K photometric
bands. \citet{lebourlot:95} have shown that this reaction rate coefficient
can be readily expressed in terms of the dust grain properties, i.e.
density $\rho_{gr}$, dust to gas mass ratio $G$, and the minimum
($a_{min}$) and maximum ($a_{max}$) values of the grain radii, for
which the size distribution is a power law $dn_{gr}\,=A_{gr}\, a^{-3.5}\, da$
~\citep{mathis:77}, subsequently quoted as the MRN distribution.
\citet{lebourlot:95} have derived the normalization factor $A_{gr}=\frac{3}{4}\,\frac{1.4\, m_{\mathrm{H}}\, G}{\pi\,\rho_{gr}}\times\frac{1}{2\left(\sqrt{a_{max}}-\sqrt{a_{min}}\right)}\times n_{\mathrm{H}}$
and the formation rate of $\mathrm{H}_{2}$:
\[
R(\mathrm{H}_{2})=\frac{1}{2}\, s\,\frac{3\times1.4\, m_{\mathrm{H}}\, G}{4\,\rho_{gr}\,\sqrt{a_{min}\, a_{max}}}\times\sqrt{\frac{8k}{\pi\, m_{\mathrm{H}}}}\times\sqrt{T}\, n(\mathrm{H})\, n_{\mathrm{H}}
\]
This expression involves a sticking coefficient $s$ for impinging
hydrogen atoms, the mass of atomic hydrogen $m_{\mathrm{H}}$, the
Boltzmann constant $k$, the gas temperature $T$. For typical dust
properties (Table~\ref{tab:Dust-properties.}), this expression becomes
$1.4\times\,10^{-17}\, s\,\sqrt{T}\, n(\mathrm{H})\, n_{\mathrm{H}}$.
Assuming that the sticking factor is $1$ up to $10\,\mathrm{K}$
and then decreases with temperature as $1/\sqrt{T}$, the formation
rate becomes equal to $4.4\times10^{-17}\, n_{\mathrm{H}}\, n(\mathrm{H})\,\mathrm{cm}^{3}\,\mathrm{s}^{-1}$.
Note that if we take a single radius of grains with a value of $0.1\,\mu\mathrm{m}$,
the formation rate is now $1.3\times10^{-17}\, n_{\mathrm{H}}\, n(\mathrm{H})\,\mathrm{cm}^{3}\,\mathrm{s}^{-1}$.\textbf{
}The present prescription also allows to derive the mean cross section
per H atom: 
\[
<n_{g}\sigma>=\frac{3}{4}\cdot\frac{1.4\, m_{\mathrm{H}}G}{\rho_{gr}}\cdot\frac{1}{\sqrt{a_{min}a_{max}}}\cdot n_{\mathrm{H}}
\]
For the standard parameters, the corresponding value is $1.95\,10^{-21}\, n_{\mathrm{H}}$.\textbf{ }

\begin{table*}
\caption{Dust properties. We assume that adsorption sites are regularly dispatched
on the grain surface so that the mean distance between them, $d$,
is the inverse of the square root of their surface number density
$n_{s}$. See references in text.\label{tab:Dust-properties.}}

\centering%
\begin{tabular}{>{\raggedright}m{4.5cm}>{\centering}m{1.8cm}>{\centering}m{1.8cm}>{\centering}m{1.8cm}>{\centering}m{1.8cm}}
\hline 
\hline Properties & ``standard'' & \citet{hasegawa:92} & amorphous carbon & Olivine\tabularnewline
\hline 
$s$ sticking factor ($1-10\,\mathrm{K}$) & $1$ &  &  & \tabularnewline
$a_{min}$ ($\mu$m) & $0.03$ &  &  & \tabularnewline
$a_{max}$ ($\mu$m) & $0.3$ &  &  & \tabularnewline
dust to gas mass ratio, $G$ & $0.01$ &  &  & \tabularnewline
density, $\rho_{gr}$ ($\mathrm{g}\,\mathrm{cm}^{-3}$)  & $3$ &  & $2.16$ & $3$\tabularnewline
$\mathrm{H}$ binding energy, $E_{b}$ ($\mathrm{meV}$)  & - & 30.16 & $56.7$ & $32.1$\tabularnewline
$\mathrm{H}$ diffusion barrier, $E_{d}$ ($\mathrm{meV}$) & - & 8.62 & $44$ & $24.7$\tabularnewline
surface density of adsorption sites, $n_{s}$ ($\mathrm{cm}^{-2}$) &  & $1.5\,10^{15}$ & $5\,10^{13}$ & $2\,10^{14}$\tabularnewline
mean distance between adsorption sites, $d_{s}\,(\textrm{\AA)}$ & $2.6$ & $2.6$ & $14.14$ & $7.07$\tabularnewline
hoping rate, $\nu_{0}$ ($s^{-1}$) with\textbf{ $d_{0}=1\,\textrm{\AA}$} & $10^{12}$ & $7.6\,10^{12}$ & $10.5\,10^{12}$ & $7.9\,10^{12}$\tabularnewline
\hline 
\end{tabular}
\end{table*}

Whereas such a simple physical picture reflects the coherence between
the observations and the basic properties of dust, a new step arose
when the quantitative values of the desorption and diffusion barriers
have been obtained from temperature desorption experiments reported
by \citet{katz:99,biham:01,biham:02,lipshtat:03,lipshtat:04} with
various types of surfaces. The corresponding values are reported in
Table~\ref{tab:Dust-properties.} and compared to older values from
\citet{hasegawa:92}. These values have been included in theoretical
models of $\mathrm{H}_{2}$ formation by \citet{biham:02} who also
discuss the relevance of using the rate equations formalism within
the LH diffusion mechanism. In the case of small grains, \citet{biham:02}
argue that only a few hydrogen atoms can stick on the surface so that
master equations describing the various probabilities of having $N$
adsorbed $\mathrm{H}$ atoms are required. This formalism leads to
a large number of coupled equations. \citet{biham:02} further introduced
the use of moment equations, reducing the number of equations and
leading, in the case of $\mathrm{H_{2}}$ formation, to only two coupled
equations. Such a procedure compares very satisfactorily to the master
equation solution \citep{biham:05}. During the diffusion process,
physisorbed $\mathrm{H}$ atoms explore the available adsorption sites
with a sweeping rate $A=\nu_{0}\,\frac{1}{4\pi a^{2}n_{s}}\,\exp\left(-\frac{T_{d}}{T_{gr}}\right)$
with $\nu_{0}$ a typical vibration frequency and $T_{d}=E_{d}/k$
the diffusion energy threshold in Kelvin. The number of available
sites, $4\pi a^{2}n_{s}$,  depends on the radius of the grain.\textbf{
}The sweeping rate probability has subsequently been revisited by
\citet{lohmar06} and\textbf{ }\citet{lohmar09} for various surface
geometries.\textbf{ }The desorption probability is assumed to follow
a similar exponential decrease $W=\nu_{0}\,\exp\left(-\frac{T_{b}}{T_{gr}}\right)$
with the same $\nu_{0}$ frequency but involves the binding energy
in Kelvin $T_{b}=E_{b}/k$. 

The frequency $\nu_{0}$ is often assumed to be equal to $10^{12}\, s^{-1}$
\citep{biham:01}. \citet{hasegawa:92} have introduced the vibrational
frequency corresponding to an harmonic oscillator within the binding
well potential in the form \textbf{$\nu{}_{0}=\frac{\sqrt{n_{s}}}{\pi}\,\sqrt{\frac{2E_{b}}{m_{\mathrm{H}}}}$.
}The usual expression is $\nu_{0}=\frac{1}{2\pi}\,\sqrt{\frac{k}{m_{\mathrm{H}}}}$,
where the force constant $k=\frac{\partial^{2}U(r)}{\partial r^{2}}\sim\frac{8\, E_{b}}{d_{0}^{2}}$
at the minimum of the interaction potential $U(r)$, which defines
a typical size of the bottom of the well $d_{0}$ by identification
with \citet{hasegawa:92} expression. \citet{kim:11} and \citet{sakong:10}
show that the stretching vibration wavenumber of $\mathrm{H}$ on
graphene is about $2600\,\mathrm{cm}^{-1}$, and the bending about
$1200\,\mathrm{cm}^{-1}$. This translates to vibration frequencies
of a few $10^{13}\,\mathrm{s}^{-1}$, rather higher than usual estimates.
We deduce a typical size $d_{0}\sim1\,\textrm{\AA}$.\textbf{ }The
corresponding values of\textbf{ $\nu_{0}=\frac{1}{\pi}\,\sqrt{\frac{2\, E_{b}}{d_{0}^{2}\, m_{\mathrm{H}}}}$
}are given in Table~\ref{tab:Dust-properties.} for various dust
compositions.\textbf{ }We have verified that model results are almost
insensitive to the value of \textbf{$\nu_{0}$.}

\citet{lepetit:09} have included the possibility of rejection of
the impinging hydrogens, when they land on an already occupied site
and have introduced these mechanisms in the Meudon PDR code together
with the size distribution of grains, but assuming a uniform dust
temperature for all grain sizes. 

The net result, for $\mathrm{H}_{2}$ formation, is that the reaction
is efficient only in a narrow window of dust temperatures, ranging
typically from $11\,\mathrm{K}$ to $19\,\mathrm{K}$, depending on
the values of desorption and diffusion barriers, as also pointed out
by \citet{chang:06}. Such a restricted range of grain temperatures
is difficult to reconcile with $\mathrm{H}_{2}$ observations towards
dense PDRs \citep{habart:04,habart:05,cazaux:04,cazaux:10,habart:11}
where the dust temperatures reach values close to and even larger
than $30\,\mathrm{K}$. As a result, new scenario have been suggested
such as the reaction between a physisorbed and a chemisorbed atom
\citep{habart:04} or reactions between two chemisorbed atoms \citep{cazaux:04,cazaux:10}. 

Other attempts to improve the physics of the actual process involve
morphological aspects of the dust grain surfaces such as roughness
effects probed through continuous time random walk Monte Carlo simulation
studies, resulting in effective formation rate coefficients depending
on the surface temperature and impinging atomic flux\citep{chang:06}.
In summary, despite the amount of laboratory and theoretical studies
done on this subject, a detailed description of $\mathrm{H_{2}}$
formation is still missing. As a consequence, several authors \citep{hollenbach:09,islam:10,kaufmann:06,sheffer:11}
prefer to use some ad hoc ``standard'' value of $3\times10^{-17}\, n_{\mathrm{H}}\, n(\mathrm{H})\,\mathrm{cm}^{3}\,\mathrm{s}^{-1}$
,\textbf{ }determined by Copernicus and FUSE observations for diffuse
clouds,\textbf{ }and / or introduce an additional multiplication factor
of 2 for dense and bright PDRs.

\subsection{$\mathrm{H}_{2}$ excitation in the formation process}

The excitation state of $\mathrm{H}_{2}$ in the formation process
is another concern. Recent experimental studies \citep{hornekaer:03,islam:07,latimer:08,lemaire:10}
have shown that $\mathrm{H}_{2}$ can be rovibrationally excited in
the formation process, where the experiments have been conducted with
two beams of $\mathrm{H}$ and $\mathrm{D}$ and various types of
surfaces. These experiments are performed on cold surfaces, indicating
that the formation mechanism involves mainly physisorbed atoms, so
a LH mechanism. Alternatively, \citet{farebrother:00,bachellerie:09,sizun:10}
have performed theoretical calculations for the ER formation mechanism
of $\mathrm{H}_{2}$ on graphenic surfaces and have derived the rovibrational
excitation of the nascent $\mathrm{H}_{2}$ molecule for different
energies of the impinging hydrogen atoms. The latest study \citep{sizun:10}
includes explicitly the role of the zero point energy, which leads
to a slight increase of the rotational excitation within $v=5$ and
$6$. In the astrophysical community, \citet{duley:93} have proposed
that infrared fluorescence of $\mathrm{H}_{2}$ could be detected
towards dark clouds, as a signature of its formation process. \citet{lebourlot:95b}
showed that the excitation due to the secondary electrons generated
by cosmic rays was in fact much more efficient, even for the standard
value of the cosmic ionization rate. Similar conclusions have recently
been derived by \citet{islam:10} in a detailed modeling study where
the most recent experimental information on the excitation of $\mathrm{H}_{2}$
has been included, assuming however the ``standard'' formation rate
of $3\times10^{-17}\, n_{H}\, n(\mathrm{H})\,\mathrm{cm}^{3}\,\mathrm{s}^{-1}$.
\citet{congiu:09} also find that internal excitation is quenched
before desorption from experiments involving amorphous water ice.
However, formation pumping signatures can also be searched for in
PDRs and were reported as such by \citet{burton:02} in the Messier
17 PDR through the mapping of the 6-4 O(3) infrared transition of
$\mathrm{H}_{2}$ at $1.733\,\mu\mathrm{m}$. We investigate the possible
consequences in the present study.

\section{\label{sec:Model-calculations}Model calculations}

\citet{lepetit:09} have studied the efficiency of $\mathrm{H}_{2}$
formation for different dust temperatures when $\mathrm{H}_{2}$ formation
takes place on amorphous carbon, assuming that the grains have the
same temperature, whatever their size. Such an hypothesis is inadequate
as discussed by different authors \citep{cuppen:06}. \citet{compiegne:11}
explicitly compute the dust temperatures resulting from the balance
between photoelectric effects and radiative infrared emission for
the different grain sizes. The results are very close to the analytic
expressions given by \citet{hollenbach:91} and used in the Meudon
PDR code \citep{lepetit:06}. We describe now two major extensions,
i.e. inclusion of the different dust grain temperatures for surface
reactions involving physisorbed hydrogen atoms and the introduction
of the ER mechanism for $\mathrm{H}_{2}$ formation. Other recent
updates of the PDR code are described in Appendix \ref{sec:Other-recent-updates}.

\subsection{Langmuir-Hinshelwood mechanism}

In a pure diffusion process (LH mechanism), physisorbed hydrogen atoms,
labelled as ``$\mathrm{H:}$'', can either diffuse on the surface,
encounter another physisorbed hydrogen atom to form $\mathrm{H}_{2}$
or leave the surface through thermal desorption or another desorption
process such as photodesorption. We derive the detailed rate equations
in App~\ref{App_Langmuir}, taking into account size dependent dust
temperatures and rejection effects. The formation rate of $\mathrm{H}_{2}$
is no longer analytic and involves a numerical integration over the
dust grain sizes. In this paper, we assume a MRN distribution law,
but the expressions are given for any distribution. The number of
particles $\mathrm{X}$ physisorbed on the surface of a grain of size
$a$ is $N_{\mathrm{X:}}(a)$ and is one of the unknowns we have to
compute.

Considering only the balance between $\mathrm{H}$ and $\mathrm{H}_{2}$
(without deuterium here), we find from App~\ref{App_Langmuir}, Eqs~(\ref{eq_adsor})
and (\ref{eq_eject}):
\[
\frac{d[\mathrm{H}]}{dt}=-k_{rej}\,\frac{S_{gr}}{d_{s}^{2}}\,[\mathrm{H}]+k_{rej}\,[\mathrm{H}]\,\int_{a_{min}}^{a_{max}}\,\left(N_{\mathrm{H:}}(a)+N_{\mathrm{H}_{2}:}(a)\right)\, dn_{g}
\]
\[
+\int_{a_{min}}^{a_{max}}\,\, k_{ev}(a)\, N_{\mathrm{H:}}(a)\, dn_{g}+\left(k_{ph}+k_{CR}\right)\,\int_{a_{min}}^{a_{max}}\, N_{\mathrm{H:}}(a)\, dn_{g}
\]

In this expression, $k_{rej}$, $k_{ev}$, $k_{ph}$ and $k_{CR}$
are the chemical reaction rate constants holding respectively for
rejection, thermal evaporation, photo-desorption and cosmic rays induced
desorption. $k_{rej}$ is proportional to the thermal velocity of
the impinging hydrogen atoms and to the sticking coefficient which
is also a function of the gas temperature. $S_{gr}$, the total grain
surface per cubic centimeter, is equal to $3\,\cdot\,\frac{1.4\, m_{\mathrm{H}}\, G}{\rho_{gr}}\,\cdot\,\frac{1}{\sqrt{a_{min}a_{max}}}\,\cdot\, n_{\mathrm{H}}$,
with the MRN distribution law, and $d_{s}$ is the mean distance between
adsorption sites\textbf{.} Among the ejection processes, only thermal
evaporation ($k_{ev}(a)$) depends on the grain size. Formation of
gas phase $\mathrm{H}_{2}$ is assumed to occur directly after the
encounter of adsorbed hydrogen atoms. The corresponding rate is given
by Eq~(\ref{eq_formLH}):
\[
\frac{d[\mathrm{H}_{2}]}{dt}=\frac{d_{s}^{2}}{4\pi}\,\int_{a_{min}}^{a_{max}}\,\frac{1}{t_{\mathrm{H}}(a)}\,\frac{1}{a^{2}}\, N_{\mathrm{H:}}^{2}(a)\, dn_{g}
\]

The hoping time $t_{\mathrm{H}}$ is dependent both on diffusion and
possible tunneling. The diffusion time, $t_{D}$, equal to $\nu_{0}^{-1}\,\exp\left(T_{d}/T_{g}(a)\right)$,
depends now on the size of the grain. The tunneling time is derived
for a rectangular barrier of height $E_{b}$ and width $d_{s}$. %
\footnote{\citet{hasegawa:93} assume a $1\,\textrm{\AA}$ width between the
two physisorbed sites, much smaller than the mean distance computed
from the density of sites displayed in Table~\ref{tab:Dust-properties.}.%
}: $t_{T}=\nu_{0}^{-1}\,\times\,\left(1+\frac{E_{b}^{2}\sinh^{2}\left(d_{s}/\lambda_{D}\right)}{4\, kT_{gr}(a)\,\left(E_{b}-kT_{g}(a)\right)}\right)$.
$\lambda_{D}$ is the de Broglie length, given by $\lambda_{D}=\hbar/\sqrt{2m_{\mathrm{H}}\left(E_{b}-kT_{g}(a)\right)}$.
The total hoping frequency is $t_{\mathrm{H}}^{-1}=t_{D}^{-1}+t_{T}^{-1}$,
is a function of the grain size.

To include these equations in the chemical scheme, the mean number
of $\mathrm{H}$ or $\mathrm{H}_{2}$ physisorbed on a grain of size
$a$, namely $N_{\mathrm{H:}}(a)$ and $N_{\mathrm{H}_{2}:}(a)$ have
to be computed. We obtain from Eq~(\ref{eq_surf_ads}), (\ref{eq_surf_ej})
and (\ref{eq_surf_rxn}):
\[
\frac{dN_{\mathrm{H:}}(a)}{dt}=k_{ad}(a)\,[\mathrm{H}]-k_{rej}\,[\mathrm{H}]\,\left(N_{\mathrm{H:}}(a)+N_{\mathrm{H}_{2}:}(a)\right)
\]
\[
-\left(k_{ev}(a)+k_{ph}+k_{CR}\right)\, N_{\mathrm{H:}}(a)-\frac{1}{t_{\mathrm{H}}(a)}\,\frac{d^{2}}{4\pi\, a^{2}}\, N_{\mathrm{H:}}^{2}(a)
\]
\[
\frac{dN_{\mathrm{H}_{2}:}(a)}{dt}=k_{ad}(a)\,[\mathrm{H}_{2}]-k_{rej}\,[\mathrm{H}_{2}]\,\left(N_{\mathrm{H:}}(a)+N_{\mathrm{H}_{2}:}(a)\right)
\]
\[
-\left(k_{ev}(a)+k_{ph}+k_{CR}\right)\, N_{\mathrm{H}_{2}:}(a)
\]
where $k_{ad}$ is the chemical rate constant for adsorption. After
discretization of the grain size distribution, the equations can be
solved as ordinary rate equations and are thus easy to incorporate
within the other chemical rate equations. However, the number of variables
grows significantly.

\subsection{Eley-Rideal mechanism\label{sub:ER}}

We explicitly introduce in the PDR model adsorption of $\mathrm{H}$
atoms into chemisorbed sites, noted as $\mathrm{H\!\!::}$, as well
as the subsequent reactions with impinging gas phase atomic hydrogen
leading to $\mathrm{\mathrm{H_{2}}}$ formation. This process has
been studied theoretically by different authors, who derive the interaction
potential with the surface and the reaction probabilities. The static
properties, potential well, possible barrier height, are significantly
dependent on the nature of the surface, which allows some liberty
on the choice of the actual value to be used in the modeling.

There is a significant difference with the adsorption process described
previously, as trapping a gas phase hydrogen into a chemisorption
site implies the crossing of a barrier where the exponential term
involves now the gas temperature. Given the variety of grain types
and surfaces properties found in the ISM, it is probable that this
barrier is not unique. Those properties are poorly known and trying
to derive a detailed model would, arbitrarily, increase the number
of free parameters. Therefore we limit ourselves to a simple approximation
that takes into account the major aspects of this process.

With the assumption that $\mathrm{H}$ is removed from a chemisorption
site only by the formation of a molecule and using results from App~\ref{App_ER},
Eq\_(\ref{eq:_ER_rate}), $\mathrm{H}_{2}$ formation rate becomes:
\begin{equation}
\left.\frac{d[\mathrm{H}_{2}]}{dt}\right|_{ER}=v_{th}\,<n\sigma_{gr}>\,\frac{\alpha(T)\,\exp\left(-\frac{T_{1}}{T}\right)}{1+\alpha(T)\,\exp\left(-\frac{T_{1}}{T}\right)}\,[\mathrm{H}]=k_{ER}\,[\mathrm{H}]\, n_{\mathrm{H}}\label{eq:k_ER}
\end{equation}

$\alpha(T)$ is an efficiency factor that depends on the gas temperature
$T$ and $<\! n\sigma_{gr}\!>$ is the total surface of grains available
per cubic centimeter (see the appendix for discussion on the parameters).
If a more complicated set of species and reactions is used (e.g.,
formation of $\mathrm{HD}$), then the equations given allow the computations
of all necessary abundances, but the rate is no longer analytic.

The corresponding rate, $k_{ER}=2.8\times10^{-17}\times\sqrt{T}\,\times\frac{\alpha(T)\,\exp\left(-T_{1}/T\right)}{\left(1+\alpha(T)\,\exp\left(-T_{1}/T\right)\right)}$
with the standard values displayed in Table~\ref{tab:Dust-properties.}.
We notice that the numerical factor is twice the value reported in
Sect.~\ref{sub:Old_form} as $\mathrm{H_{2}}$ is now produced from
the collision between two different partners, a gas phase $\mathrm{H}$
atom and a chemisorbed $\mathrm{H\!\!::}$. The order of magnitude
of this chemical rate constant is again within the ``standard''
value derived by \citet{jura:74}. The actual value of $k_{ER}$ depends
on the chemisorption barrier $T_{1}$, and on the sticking efficiency
$\alpha(T)$ but not on the distance between adsorption sites. The
influence of $T_{1}$ and $\alpha(T)$ on the formation efficiency
and line intensities is discussed in App~\ref{App_ER}.

We will see below that this reaction has a major effect on thermal
balance at the edge of the cloud as each $\mathrm{H}_{2}$ formation
as well as each destruction contributes to the gas heating. This is
thus a very efficient way to couple the strong ultraviolet radiation
field to the gas.

\subsection{Excitation during formation\label{sub:Excitation}}

$\mathrm{H}_{2}$ formation releases about $4.5\,\mathrm{eV}$. Most
current prescriptions assume an equipartition of this energy between
translational energy, $\mathrm{H}_{2}$ internal energy and heating
of the grain. In this way, about $17000\,\mathrm{K}$ is spread in
$\mathrm{H}_{2}$ rovibrational states. Other scenarios have been
proposed by \citet{black:87}. Recent experimental results obtained
on cold graphitic surfaces \citep{islam:10} or on silicates \citep{lemaire:10}
show that high vibrational levels are preferentially populated in
the formation process.

The Meudon PDR code permits to choose between several scenarios. In
this paper, for the LH mechanism, we adopt the default option in which
1/3 of the formation enthalpy is spread in internal energy with a
Boltzmann distribution among rovibrational levels (see the description
in \citealp{lepetit:06}). Note that this is physically relevant only
if enough levels of $\mathrm{H}_{2}$ are included in the computations,
even if their steady-state populations are negligible.

For the ER process, we follow the energy distribution obtained by
\citet{sizun:10} (their Table~1) and consider that the uncertainties
are larger than the differences that they computed at incident energies
of $15$ and $50\,\mathrm{meV}$ for the impinging $\mathrm{H}$ atom.
The energy available after formation of a $\mathrm{H}_{2}$ molecule
is thus split into:
\begin{itemize}
\item $0.3\,\mathrm{eV}\,(\sim3500\,\mathrm{K})$: Depth of chemical bound
potential well.
\item $2.7\,\mathrm{eV}\,(\sim31300\,\mathrm{K})$: Internal energy, spread
following a Boltzmann distribution (mean level: $v\sim5$, $J\sim6$)
\item $0.6\,\mathrm{eV}\,(\sim7000\,\mathrm{K})$: Translational kinetic
energy (heating term for the gas).
\item $1\,\mathrm{eV}\,(\sim11600\,\mathrm{K})$: Heating of the grain.
\end{itemize}
Note that this last term is not yet included in the thermal balance
of the grains.

\subsection{Heating\label{sub:Heating}}

This enhanced formation rate at the edge of PDRs provides a new contribution
to the gas heating. At the edge of clouds, since the $\mathrm{H}_{2}$
photodissociation rate is important and the ER mechanism is efficient,
the formation-destruction cycle of $\mathrm{H}_{2}$ has a short time
scale. Each destruction by a UV photon leads to about $0.6\,\mathrm{eV}$
of kinetic energy, but each subsequent formation also provides a way
to tap the formation enthalpy reservoir. As seen above (Sect.~\ref{sub:Excitation}),
most of this energy is in high lying rovibrational levels. So, one
$\mathrm{H}_{2}$ formation is usually followed by emission of an
IR photon, which may be detected. However, for high enough densities,
collisional deexcitation with atomic $\mathrm{H}$ comes in competition
with spontaneous radiative decay. Above a critical density the collisional
process dominates%
\footnote{e.g., for the preferentially populated level $v=5$, $J=6$, the total
radiative decay probability is $A_{6,5}=3.7\,10^{-6}\,\mathrm{s}^{-1}$.
Collision deexcitation rates with $\mathrm{H}$ at $1000\,\mathrm{K}$
are typically a few times $10^{-12}\, cm^{3}s^{-1}$ $ $, which gives
critical densities close to $10^{6}\,\mathrm{cm}^{-3}$.%
} and the released kinetic energy provides an efficient heating mechanism.
This may lead to gas temperatures in the range of a few thousands,
which increases further the formation rate (which is proportional
to $v_{th}$, see Eq~(\ref{eq:k_ER})), decreases the cycling time
scales and enhances still further the heating rate.

This positive loop is quenched by the fact that our sticking coefficient
goes to zero with increasing temperature, which allows for an equilibrium
temperature to be reached.

\begin{figure}
\centering\includegraphics[width=1\columnwidth]{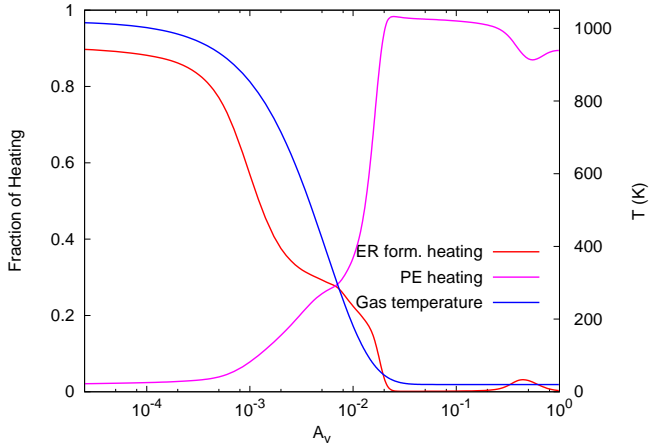}

\caption{Respective weight of ER $\mathrm{H}_{2}$ formation and photo-electric
heating in a high density PDR. Here the density is $n_{\mathrm{H}}=10^{6}\,\mathrm{cm}^{-3}$
and the radiation field $G_{0}=10^{3}$. The temperature profile is
fixed and has been explicitly chosen to illustrate the importance
of ER heating at high temperature.\label{fig:Heating}}
\end{figure}

For a high density and a high radiation field, this process is by
far the dominant one at the edge of the cloud. Fig.~\ref{fig:Heating}
shows the fraction of heating provided by ER $\mathrm{H}_{2}$ formation
vs. photo-electric effect in a cloud where the temperature profile
is fixed. Close to the edge, the high gas temperature allows for the
ER process to be fast and \textbf{most} of the heating comes from
collisional de-excitation. Once the temperature drops, that process
is quenched and photo-electric heating dominates as long as the optical
depth is not too high.

The transition region is chemically very active ($\mathrm{H}$ and
$\mathrm{H}_{2}$ being both abundant, and the temperature still quite
high). Formation rates of molecules such as $\mathrm{OH}$ or $\mathrm{CH}^{+}$
may be enhanced in particular if $\mathrm{H}_{2}$ internal energy
can contribute to overcome the energy thresholds of the chemical reaction.
Heat released by exothermic reactions also provides most of the heating
missing from the two processes of Fig.~\ref{fig:Heating}.

\section{Model results\label{sec:Model-results}}

In this section, we study the effect of different prescriptions for
molecular hydrogen formation on grains.\textbf{ }We model isobaric
clouds since this is the simplest equation of state for dilute gas
at steady-state embedded in a stationary environment where gravitation
is negligible: large pressure variations (such as found in a constant
density cloud with large temperature variations) would lead to some
kind of relaxation. Smaller entities (clumps, filaments, etc...) may
be better described as isochoric clouds within these large structures.
A grid of such isochoric models, with the same physics than the isobaric
models presented below has also been produced. They are not discussed
in this paper but can be found in the PDR Data Base (http://pdr.obspm.fr).

Models labelled ``A'' correspond to a fixed formation rate of molecular
hydrogen of $3\,\times\,10^{-17}\,\mathrm{cm}^{3}\,\mathrm{s}^{-1}$,
following prescriptions by \citet{hollenbach:09,islam:10}; in models
``B'', $\mathrm{H}_{2}$ is formed only by LH mechanism, in models
``C'' LH and ER mechanisms are both activated to form $\mathrm{H}_{2}$.
The model geometry is a plane parallel infinite slab with an isotropic
radiation field impinging on both sides.

We explore a large domain of gas pressure and intensity of the incident
UV radiation field. The gas pressure is given by $P=n\times T=\left[n(\mathrm{H})+n(\mathrm{H_{2})+n(\mathrm{He)}}\right]\times T$.
The radiation field is scaled versus the ISRF evaluated by \citet{mathis:83}
in the UV and visible part of the spectra. Longward of $\sim6000\,\textrm{\AA}$
the ISRF is corrected using infrared observations from \citet[F. Boulanger and M. Gonzalez, private communication]{hauser:98}.
The scaling factor $\chi$ is applied only in the VUV range, from
$13.6$ to $5.17\,\mathrm{eV}$ (corresponding to the wavelength range
$912$ to $2400\,\textrm{\AA}$) to mimic the presence of nearby bright
OB stars. It represents the scaling factor of the UV part of the radiation
field at a point that receives the radiation field from $4\pi$ steradians. 

\begin{table}
\caption{Grains parameters and elemental abundances. $\mathrm{Fe}$ stands
for a generic metallic ion.\label{tab:ParametreModeles}}
\centering%
\begin{tabular}{lccc}
\hline 
\hline Parameter & Definition & Value & Unit\tabularnewline
\hline 
$\omega$ & dust albedo & $0.42$ & -\tabularnewline
$g$ & anisotropy factor & $0.6$ & -\tabularnewline
$R_{V}$ & $A_{V}/E(B-V)$ & $3.1$ & -\tabularnewline
$C_{D}$ & $N_{\mathrm{H}}/E(B-V)$ & $5.8\times10^{21}$ & $\mathrm{cm}^{-2}\,\mathrm{mag}^{-1}$\tabularnewline
$\zeta$ & cosmic rays flux & $5\times10^{-17}$ & $/s\,/\mathrm{H_{2}}$\tabularnewline
$v_{turb}$ & turbulent velocity & $3$ & $\mathrm{km}\,\mathrm{s}{}^{-1}$\tabularnewline
$\delta_{\mathrm{He}}$ & $\{\mathrm{He\}}/\{\mathrm{H\}}$ & 0.1 & -\tabularnewline
$\delta_{\mathrm{O}}$ & $\{\mathrm{O\}}/\{\mathrm{H\}}$ & $3.19\times10^{-4}$ & -\tabularnewline
$\delta_{\mathrm{C}}$ & $\{\mathrm{C\}}/\{\mathrm{H\}}$ & $1.32\times10^{-4}$ & -\tabularnewline
$\delta_{\mathrm{N}}$ & $\{\mathrm{N\}}/\{\mathrm{H\}}$ & $7.5\times10^{-5}$ & -\tabularnewline
$\delta_{\mathrm{S}}$ & $\{\mathrm{S\}}/\{\mathrm{H\}}$ & $1.86\times10^{-5}$ & -\tabularnewline
$\delta_{\mathrm{Fe}}$ & $\{\mathrm{Fe\}}/\{\mathrm{H\}}$ & $1.5\times10^{-8}$ & -\tabularnewline
\hline 
\end{tabular}
\end{table}

Dust properties, elemental abundances (\{X\} stands for the total
abundance of the X element) and other parameters are given in Tables~\ref{tab:Dust-properties.}
and \ref{tab:ParametreModeles}. The adopted dust extinction curve
corresponds to the parameters of \citet{fitzpatrick:90}for the Galaxy.
Photodissociation probabilities of $\mathrm{H_{2}}$ and $\mathrm{CO}$,
are explicitly computed from the different discrete photodissociating
transitions. We include a turbulent component to the thermal line
width and take a common value of $3\,\mathrm{km}\,\mathrm{s}^{-1}$
for all transitions. We choose to display results of isobaric models
with values relevant both to diffuse and dense gas.

\subsection{Diffuse gas\label{sub:Diffuse-gas}}

The chemical network includes only gas phase processes, except for
$\mathrm{H}$ and $\mathrm{H_{2}}$ as well as neutralization of atomic
ions on grain surfaces. It involves $131$ chemical species and about
$2650$ chemical reactions. The analysis of UV absorption of neutral
carbon lines towards a sample of translucent and diffuse line of sights
\citep{jenkins:79,jenkins:01,jenkins:07} allows to derive the range
of thermal pressures relevant for the galactic interstellar gas. Table~\ref{tab_diffus}
displays results for two isobaric chemical models corresponding to
a total visual magnitude $A_{V}^{tot}=1$ and $\chi=1$ on both sides
of the cloud. The impinging flux at the edge is about $0.56$ times
the Mathis field as some photons are coming from the opposite side
of the cloud or are back-scattered by dust. The corresponding radiation
pressure at the edge is $u/3$, where $u$ is the energy density per
unit volume, i.e. $1.7\,10^{-14}\,\mathrm{dynes\, cm^{-2}}$, always
smaller than the considered thermal pressures (respectively $4.14\,10^{-13}$
and $1.38\,10^{-12}\,\mathrm{dynes\, cm^{-2}}$). The size in $\mathrm{pc}$
is noted $l$ (see conversion in App~\ref{sec:Size_conv}). The grains
temperatures depend only on the radiation field intensity. Their common
values at the edge of the clouds span a range between $10.7$ and
$12.5\,\mathrm{\mathrm{K}}$.

\begin{table*}
\caption{Model results for diffuse clouds conditions.\textbf{ }Models A, B
and C are defined in the text. Figures in parentheses correspond to
powers of ten. Molecular fraction, $f,$ is defined as $2\, N(\mathrm{H}_{2})/[2\, N(\mathrm{H}_{2})+N(\mathrm{H})]$.
$T_{01}$ is the excitation temperature derived from the ratio of
the column densities of the first levels of molecular hydrogen.\textbf{
}Superscript ``obs'' is for ``edge of the cloud on the observer
side'' and ``center'' is for ``center of the cloud''.\textbf{
}Column densities $N(\mathrm{X})$ are in $\mathrm{cm}^{-2}$. \label{tab_diffus}}

\centering%
\begin{tabular}{>{\raggedright}p{2cm}cccccc}
\hline 
\hline Pressure & \multicolumn{3}{c}{$P=3\,10^{3}\,(\mathrm{cm}^{-3}\,\mathrm{K})$} & \multicolumn{3}{c}{$P=10^{4}\,(\mathrm{cm}^{-3}\,\mathrm{K})$}\tabularnewline
Model & A & B & C & A & B  & C\tabularnewline
\hline 
$l$ $(\mathrm{pc})$ & 14.6 & 14.7 & 14.7 & 1.5 & 1.5 & 1.5\tabularnewline
$n_{\mathrm{H}}^{obs}$$(\mathrm{cm}^{-3})$ & 3.69 & 2.41 & 1.17 & 186 & 166 & 165\tabularnewline
$T^{obs}$$(\mathrm{K})$ & 739 & 1128 & 2321 & 49 & 55 & 55\tabularnewline
$R_{\mathrm{H_{2}}}^{obs}$$(\mathrm{cm}^{3}\,\mathrm{s}^{-1})$ & 3.0(-17) & 5.8(-17) & 1.8(-16) & 3.0(-17) & 5.8(-17) & 5.9(-17)\tabularnewline
$n_{\mathrm{H}}^{center}$ $(\mathrm{cm}^{-3})$ & 74 & 73 & 73 & 543 & 543 & 543\tabularnewline
$T^{center}$$(\mathrm{K})$ & 65 & 67 & 67 & 31 & 31 & 31\tabularnewline
$N(\mathrm{H})$ & 5.1(20) & 3.4(20) & 2.6(20) & 8.0(19) & 4.6(19) & 4.5(19)\tabularnewline
$N(\mathrm{H_{2}})$ & 6.8(20) & 7.7(20) & 8.1(20) & 9.0(20) & 9.1(20) & 9.1(20)\tabularnewline
$f$ & 0.72 & 0.82 & 0.86 & 0.96 & 0.97 & 0.98\tabularnewline
$T_{01}$ $(\mathrm{K})$ & 73 & 78 & 80 & 36 & 37 & 37\tabularnewline
$N(\mathrm{C^{+}})$ & 2.4(17) & 2.5(17) & 2.5(17) & 2.2(17) & 2.2(17) & 2.2(17)\tabularnewline
$N(\mathrm{C})$ & 2.0(15) & 1.9(15) & 1.9(15) & 2.8(16) & 2.8(16) & 2.8(16)\tabularnewline
$N(\mathrm{O})$ & 6.0(17) & 6.0(17) & 6.0(17) & 6.0(17) & 6.0(17) & 6.0(17)\tabularnewline
$N(\mathrm{CO})$ & 1.4(13) & 1.2(13) & 1.2(13) & 5.8(13) & 7.0(13) & 7.0(13)\tabularnewline
$N(\mathrm{OH})$ & 3.2(13) & 3.1(13) & 3.0(13) & 2.7(12) & 2.9(12) & 2.9(12)\tabularnewline
$N(\mathrm{CH})$ & 1.1(12) & 1.4(12) & 1.4(12) & 1.6(13) & 1.8(13) & 1.8(13)\tabularnewline
$N(\mathrm{NH})$ & 3.8(10) & 4.1(10) & 4.3(10) & 2.8(10) & 2.8(10) & 2.8(10)\tabularnewline
$N(\mathrm{CN})$ & 2.0(10) & 2.3(10) & 2.4(10) & 1.4(12) & 1.6(12) & 1.6(12)\tabularnewline
$N(\mathrm{OH^{+}})$ & 2.3(12) & 1.9(12) & 1.8(12) & 9.9(9) & 1.1(10) & 1.1(10)\tabularnewline
$N(\mathrm{H_{2}O^{+}})$ & 9.6(11) & 9.3(11) & 8.8(11) & 9.9(9) & 1.1(10) & 1.1(10)\tabularnewline
$N(\mathrm{H_{3}^{+}})$ & 2.0(13) & 2.8(13) & 3.1(13) & 3.4(12) & 3.7(12) & 3.7(12)\tabularnewline
\hline 
\end{tabular}
\end{table*}

We note that the results are typical of cold neutral medium (CNM)
properties, following \citet{wolfire:95}. We also find that the dependence
of the displayed quantities are marginally dependent on the assumed
formation scenario of molecular hydrogen. This results directly from
the range of gas and dust temperatures and the assumed dust size distribution.
Thus, in diffuse clouds, the role of the ER mechanism is negligible
except at the edge of the low pressure models ($P=3\,10^{3}\,(\mathrm{cm}^{-3}\,\mathrm{K})$
= $4.14\,10^{-13}$ $\mathrm{dynes\, cm^{-2}}$). We also report $T_{01}$,
the excitation temperature derived from the computed column densities
of $\mathrm{H_{2}}$, J = 0 and J = 1 levels: $T_{01}\,=\,170.5/Log(9N_{0}/N_{1})$.
The values are in agreement with the observations ($T\simeq67\mathrm{K}$
from FUSE observations, \citet{burgh:07}, \citealp{rachford09},
\citet{Sheffer:08}), especially for models with $P=3\,10^{3}\,\mathrm{cm^{-3}\, K}$.
For the low pressure models, we note that $\mathrm{H}_{2}$ formation
is enhanced in models B and C compared to model A leading to a significantly
higher value of the molecular fraction. The FUSE survey of diffuse
clouds \citep{rachford09} has not detected clouds with molecular
fraction over 0.6. This is one of the difficulties in reconciling
models of diffuse clouds and observations. First, molecular fractions
determined by observations also include the atomic hydrogen content
over the full line of sight. Second, the molecular fraction depends
strongly on the size of the clouds and on the UV illumination. The
detailed study is beyond the scope of this paper.

\subsection{Dense PDRs\label{sub:Dense-PDRs}}

Dense and bright PDRs offer the opportunity to detect infrared transitions
of molecular hydrogen and fine structure transitions of neutral and
ionized atoms ($\mathrm{O}$, $\mathrm{C}$, $\mathrm{C^{+}}$ etc..)
which are then used to derive the relevant physical conditions by
the observers. We again consider isobaric models irradiated on the
observer side by 3 different radiation fields, respectively $10^{2}$,
$10^{3}$ and $10^{4}$ times the standard interstellar radiation
field of \citet{mathis:83}. The back side side is illuminated by
standard radiation field ($\chi=1$). A total visual magnitude of
$10$ is assumed.

\subsubsection{$\mathrm{H_{2}}$ formation \label{sub:H2-formation}}

We first check the role of the formation process in the atomic to
molecular transition properties. Tables~\ref{tab:p5}, \ref{tab:p6},
\ref{tab:p7} and \ref{tab:p8} display results for 4 different pressures
and 3 radiation field enhancement factors.

\begin{table*}
\caption{Model results for $P=10^{5}\,\mathrm{cm}^{-3}\,\mathrm{K}$ and 3
different radiation field enhancements.\textbf{ }Models A, B and C
are identical as previously and defined in the text. $l$ is the total
width of the cloud expressed in $\mathrm{pc}$, corresponding to a
total visual magnitude of $10$. $N(\mathrm{X})$ stands for the resulting
total column density of species $\mathrm{X}$. Exponent ``obs''
means values at the edge of the cloud on the observer side. Numbers
in parenthesis give the powers of ten.\label{tab:p5}}
\centering%
\begin{tabular}{>{\raggedright}p{2.6cm}ccccccccc}
\hline 
\hline $\chi^{obs}$ & \multicolumn{3}{c}{$10^{2}$} & \multicolumn{3}{c}{$10^{3}$} & \multicolumn{3}{c}{$10^{4}$}\tabularnewline
$T_{g}^{obs}$ (min) ($\mathrm{K}$) & \multicolumn{3}{c}{16.7} & \multicolumn{3}{c}{26.2} & \multicolumn{3}{c}{41}\tabularnewline
$T_{g}^{obs}$ (max) ($\mathrm{K}$) & \multicolumn{3}{c}{27.7} & \multicolumn{3}{c}{44.1} & \multicolumn{3}{c}{70}\tabularnewline
Model & A & B & C & A & B & C & A & B & C\tabularnewline
\hline 
$l$ ($\mathrm{pc}$) & 1.0 & 1.0 & 1.0 & 2.1 & 2.2 & 2.1 & 3.15 & 3.6 & 3.4\tabularnewline
$n_{\mathrm{H}}^{obs}$ ($\mathrm{cm}^{-3}$) & 364 & 374 & 353 & 398 & 414 & 377 & 566 & 584 & 566\tabularnewline
 $T^{obs}$ ($\mathrm{K}$) & 250 & 243 & 258 & 228 & 219 & 241 & 161 & 156 & 160\tabularnewline
$R_{\mathrm{H_{2}}}^{obs}$ ($\mathrm{cm}^{3}\,\mathrm{s}^{-1}$) & 3(-17) & 1.2(-18) & 1.1(-16) & 3(-17) & 1.9(-26) & 9.8(-17) & 3.0(-17) & 5.2(-31) & 5.3(-17)\tabularnewline
$A_{V}${\footnotesize{} ($\mathrm{H}=\mathrm{H}_{2}$)} & 0.41 & 0.55 & 0.28 & 0.94 & 1.5 & 0.78 & 1.54 & ... & 3.6\tabularnewline
$n_{\mathrm{H}\,(\mathrm{H=H_{2})}}$ ($\mathrm{cm}^{-3}$) & 930 & 1.3(3) & 747 & 750 & 1.58(3) & 676 & 706 & ... & 3.6(3)\tabularnewline
$T_{(\mathit{\mathrm{H=H_{2})}}}$ ($\mathrm{K}$) & 142 & 97 & 172 & 172 & 83 & 193 & 182 & ... & 35\tabularnewline
$N(\mathrm{H})$ ($\mathrm{cm}^{-2}$) & 6.7(20) & 9.8(20) & 5.0(20) & 1.7(21) & 2.8(21) & 1.4(21) & 2.8(21) & 1.7(22) & 1.3(22)\tabularnewline
$N(\mathrm{H_{2}})$ ($\mathrm{cm}^{-2}$) & 9.0(21) & 8.9(21) & 9.1(21) & 8.5(21) & 8.0(21) & 8.6(21) & 8.0(21) & 8.1(20) & 2.9(21)\tabularnewline
\hline 
\end{tabular}
\end{table*}

\begin{table*}
\caption{Same as Table~\ref{tab:p5} for $P=10^{6}\,\mathrm{cm}^{-3}\,\mathrm{K}$.\label{tab:p6}}

\centering%
\begin{tabular}{>{\raggedright}p{2.6cm}ccccccccc}
\hline 
\hline $\chi^{obs}$ & \multicolumn{3}{c}{$10^{2}$} & \multicolumn{3}{c}{$10^{3}$} & \multicolumn{3}{c}{$10^{4}$}\tabularnewline
$T_{g}^{obs}$ (min) ($\mathrm{K}$) & \multicolumn{3}{c}{16.7} & \multicolumn{3}{c}{26.2} & \multicolumn{3}{c}{41}\tabularnewline
$T_{g}^{obs}$ (max) ($\mathrm{K}$) & \multicolumn{3}{c}{27.7} & \multicolumn{3}{c}{44.1} & \multicolumn{3}{c}{70}\tabularnewline
Model & A & B & C & A & B & C & A & B & C\tabularnewline
\hline 
$l$ ($\mathrm{pc}$) & 0.07 & 0.07 & 0.07 & 0.17 & 0.17 & 0.17 & 0.32 & 0.37 & 0.36\tabularnewline
$n_{\mathrm{H}}^{obs}$ ($\mathrm{cm}^{-3}$) & 5.9(3) & 6.4(3) & 5.9(3) & 3.1(3) & 3.4(3) & 2.7(3) & 3.6(3) & 3.9(3) & 3.3(3)\tabularnewline
 $T^{obs}$ ($\mathrm{K}$) & 154 & 141 & 154 & 289 & 269 & 339 & 250 & 235 & 272\tabularnewline
$R_{\mathrm{H_{2}}}^{obs}$ ($\mathrm{cm}^{3}\,\mathrm{s}^{-1}$) & 3.0(-17) & 1.9(-18) & 5.1(-17) & 3.0(-17) & 1.6(-25) & 1.4(-16) & 3.0(-17) & 3.4(-30) & 1.1(-16)\tabularnewline
$A_{V}${\footnotesize{} ($\mathrm{H}=\mathrm{H}_{2}$)} & 0.065 & 0.21 & 0.042 & 0.51 & 1.0 & 0.32 & 1.1 & ... & 0.85\tabularnewline
$n_{\mathrm{H}\,(\mathrm{H=H_{2})}}$ ($\mathrm{cm}^{-3}$) & 9.0(3) & 1.4(4) & 8.3(3) & 6.4(3) & 1.6(4) & 5.0(3) & 5.6(3) & ... & 4.8(3)\tabularnewline
$T_{(\mathit{\mathrm{H=H_{2})}}}$ ($\mathrm{K}$) & 143 & 90 & 158 & 205 & 83 & 257 & 234 & ... & 268\tabularnewline
$N(\mathrm{H})$ ($\mathrm{cm}^{-2}$) & 1.3(20) & 3.7(20) & 1.1(20) & 8.8(20) & 1.9(21) & 6.0(20) & 2.0(21) & 1.7(22) & 1.1(22)\tabularnewline
$N(\mathrm{H_{2}})$ ($\mathrm{cm}^{-2}$) & 9.3(21) & 9.2(21) & 9.3(21) & 8.9(21) & 8.4(21) & 9.1(21) & 8.4(21) & 1.1(21) & 3.9(21)\tabularnewline
\hline 
\end{tabular}
\end{table*}

\begin{table*}
\caption{Same as Table~\ref{tab:p5} for $P=10^{7}\,\mathrm{cm}^{-3}\,\mathrm{K}$.\label{tab:p7}}
\centering%
\begin{tabular}{>{\raggedright}p{2.6cm}ccccccccc}
\hline 
\hline $\chi^{obs}$ & \multicolumn{3}{c}{$10^{2}$} & \multicolumn{3}{c}{$10^{3}$} & \multicolumn{3}{c}{$10^{4}$}\tabularnewline
$T_{g}^{obs}$ (min) ($\mathrm{K}$) & \multicolumn{3}{c}{16.7} & \multicolumn{3}{c}{26.2} & \multicolumn{3}{c}{41}\tabularnewline
$T_{g}^{obs}$ (max) ($\mathrm{K}$) & \multicolumn{3}{c}{27.7} & \multicolumn{3}{c}{44.1} & \multicolumn{3}{c}{70}\tabularnewline
Model & A & B & C & A & B & C & A & B & C\tabularnewline
\hline 
$l$ ($\mathrm{pc}$) & 6.4(-3) & 6.4(-3) & 6.4(-3) & 1.3(-2) & 1.4(-2) & 1.3(-2) & 3.2(-2) & 3.5(-2) & 3.7(-2)\tabularnewline
$n_{H}^{obs}$ ($\mathrm{cm}^{-3}$) & 7.0(4) & 1.1(5) & 7.2(4) & 3.1(4) & 3.6(4) & 2.1(4) & 2.2(4) & 2.5(4) & 1.7(4)\tabularnewline
$T^{obs}$ ($\mathrm{K}$) & 130 & 81 & 126 & 295 & 253 & 432 & 413 & 366 & 532\tabularnewline
$R_{\mathrm{H_{2}}}^{obs}$ ($\mathrm{cm}^{3}\,\mathrm{s}^{-1}$) & 3.0(-17) & 2.8(-18) & 3.3(-17) & 3.0(-17) & 1.7(-24) & 1.6(-16) & 3.0(-17) & 2.2(-29) & 1.7(-16)\tabularnewline
$A_{V}${\footnotesize{} ($\mathrm{H}=\mathrm{H}_{2}$)} & 2.4(-3) & 3.0(-2) & 3.5(-3) & 0.16 & 0.70 & 0.066 & 0.70 & 2.8 & 0.50\tabularnewline
$n_{\mathrm{H}\,(\mathrm{H=H_{2})}}$ ($\mathrm{cm}^{-3}$) & 1.1(5) & 1.7(5) & 1.3(5) & 6.0(4) & 1.7(5) & 4.0(4) & 4.5(4) & 3.2(5) & 3.4(4)\tabularnewline
$T_{(\mathit{\mathrm{H=H_{2})}}}$ ($\mathrm{K}$) & 114 & 77 & 97 & 217 & 75 & 321 & 292 & 40 & 389\tabularnewline
$N(\mathrm{H})$ ($\mathrm{cm}^{-2}$) & 1.3(19) & 7.1(19) & 2.9(19) & 2.9(20) & 1.3(21) & 1.8(20) & 1.3(21) & 1.1(22) & 1.9(21)\tabularnewline
$N(\mathrm{H_{2}})$ ($\mathrm{cm}^{-2}$) & 9.3(21) & 9.3(21) & 9.3(21) & 9.2(21) & 8.7(21) & 9.3(21) & 8.7(21) & 3.8(21) & 8.4(21)\tabularnewline
\hline 
\end{tabular}
\end{table*}

\begin{table*}
\caption{Same as Table~\ref{tab:p5} for $P=10^{8}\,\mathrm{cm}^{-3}\,\mathrm{K}$.\label{tab:p8}}

\centering%
\begin{tabular}{>{\raggedright}p{2.6cm}ccccccccc}
\hline 
\hline $\chi^{obs}$ & \multicolumn{3}{c}{$10^{2}$} & \multicolumn{3}{c}{$10^{3}$} & \multicolumn{3}{c}{$10^{4}$}\tabularnewline
$T_{g}^{obs}$ (min) ($\mathrm{K}$) & \multicolumn{3}{c}{16.7} & \multicolumn{3}{c}{26.2} & \multicolumn{3}{c}{41}\tabularnewline
$T_{g}^{obs}$ (max) ($\mathrm{K}$) & \multicolumn{3}{c}{27.7} & \multicolumn{3}{c}{44.1} & \multicolumn{3}{c}{70}\tabularnewline
Model & A & B & C & A & B & C & A & B & C\tabularnewline
\hline 
$l$ ($\mathrm{pc}$) & 6.8(-4) & 6.1(-4) & 6.1(-4) & 1.2(-3) & 1.2(-3) & 1.3(-3) & 4.0(-3) & 3.7(-3) & 5.2(-3)\tabularnewline
$n_{\mathrm{H}}^{obs}$ ($\mathrm{cm}^{-3}$) & 2.4(5) & 7.4(5) & 6.6(4) & 1.8(5) & 5.1(5) & 5.5(4) & 9.2(4) & 1.2(5) & 5.8(4)\tabularnewline
 $T^{obs}$ ($\mathrm{K}$) & 375 & 123 & 1386 & 502 & 179 & 1647 & 990 & 773 & 1565\tabularnewline
$R_{\mathrm{H_{2}}}^{obs}$ ($\mathrm{cm}^{3}\,\mathrm{s}^{-1}$) & 3.0(-17) & 3.5(-18) & 1.3(-16) & 3.0(-17) & 2.4(-24) & 1.5(-16) & 3(-17) & 1.0(-28) & 1.5(-16)\tabularnewline
$A_{V}${\footnotesize{} ($\mathrm{H}=\mathrm{H}_{2}$)} & 7(-3) & 2.0(-3) & 4.9(-4) & 4.0(-2) & 0.50 & 1.7(-2) & 0.48 & 2.1 & 0.40\tabularnewline
$n_{\mathrm{H}\,(\mathrm{H=H_{2})}}$ ($\mathrm{cm}^{-3}$) & 4.4(5) & 1.4(6) & 2.4(5) & 4.4(5) & 1.75(6) & 2.1(5) & 3.2(5) & 2.6(6) & 2.1(5)\tabularnewline
$T_{(\mathit{\mathrm{H=H_{2})}}}$ ($\mathrm{K}$)  & 296 & 91 & 534 & 298 & 74 & 601 & 411 & 48 & 618\tabularnewline
$N(\mathrm{H})$ ($\mathrm{cm}^{-2}$) & 3.1(19) & 1.5(19) & 8.7(18) & 1.0(20) & 9.0(20) & 9.3(19) & 8.4(20) & 4.8(21) & 7.2(20)\tabularnewline
$N(\mathrm{H_{2}})$ ($\mathrm{cm}^{-2}$) & 9.3(21) & 9.3(21) & 9.4(21) & 9.3(21) & 8.9(21) & 9.3(21) & 8.9(21) & 6.9(21) & 9.0(21)\tabularnewline
\hline 
\end{tabular}
\end{table*}

The range of dust temperatures at the edge of the clouds has a dramatic
effect on the formation rate of $\mathrm{H_{2}}$, when only diffusion
mechanisms are involved (B models), as expected. In this latter case,
there is even no range of visual magnitudes where the local density
of $\mathrm{H}_{2}$ exceeds that of atomic $\mathrm{H}$, when the
pressure is $10^{6}\,\mathrm{cm}^{-3}\,\mathrm{K}$ or smaller (shown
as \textbf{an} ellipsis in the corresponding cells in Tables~\ref{tab:p5}
and \ref{tab:p6}). The different trends can be understood from analytic
developments obtained in a single grain size approximation, as given
in Apps.~\ref{App_Langmuir} and \ref{App_ER}. The effect of pressure
is best seen with the $\chi=10^{3}$ results:
\begin{itemize}
\item Grains are warm ($26$ to $44\,\mathrm{K}$), so that thermal evaporation
of physisorbed hydrogen is efficient. Therefore the ER process dominates
over the LH one.
\item All things equal, increasing pressure in case B (where only LH is
active) gives results in accordance with Eq~(\ref{eq:LH_lowH}).
Temperature at the edge does not vary much, so that $[\mathrm{H}]\propto P$
and $R_{\mathrm{H}_{2}}$ is indeed proportional to $[\mathrm{H}]$
as expected.
\item Including the ER process, $R_{\mathrm{H}_{2}}$ first rise with increasing
$P$ due to a (at first) slow rise of temperature. Then for the highest
pressure, the heating mechanism described in Sect.~\ref{sub:Heating}
is fully at play, and $R_{\mathrm{H}_{2}}$ decreases as the temperature
stabilizes on the right side of formation rate curve of Fig.~\ref{fig_ER_rate}.
\end{itemize}
Increasing $\chi$ leads to similar results, with still higher grain
temperatures. However complex UV pumping processes have an increasing
role and Eq~(\ref{eq:LH_lowH}) is less accurate. For a lower radiation
field, the range of $\mathrm{H}$ densities spanned by increasing
pressure is within the critical densities defined in \ref{App_Langmuir}
and the full Eq~(\ref{eq:LH_full}) applies. In both cases, the ER
rate follows closely the temperature behavior with a positive feedback
being most efficient at high pressure. 

We see that allowing for the possibility of chemisorption compensates,
in a natural way, the deficiencies of the LH mechanism when dust grains
become too warm. The values of the formation rate of molecular hydrogen
attained in such bright PDR conditions become then very comparable
to the values deduced from the observations \citep{habart:04} with
an increase by a factor 2 to 5 of the standard $3\,10^{-17}\,\mathrm{n_{\mathrm{H}}\, n(\mathrm{H)\,}cm^{3}\,\mathrm{s^{-1}}}$
value depending on impinging incident radiation field\textbf{.} Even
if the chemisorption properties are somewhat empirical (the dependence
is discussed in App.~\ref{App_ER}), we feel that such a mechanism
allows to solve a major difficulty in the theory of $\mathrm{H_{2}}$
formation.

\subsubsection{Emission lines\label{sub:Emission-lines}}

\begin{figure*}
\includegraphics[bb=60bp 75bp 370bp 275bp,clip,width=0.5\linewidth]{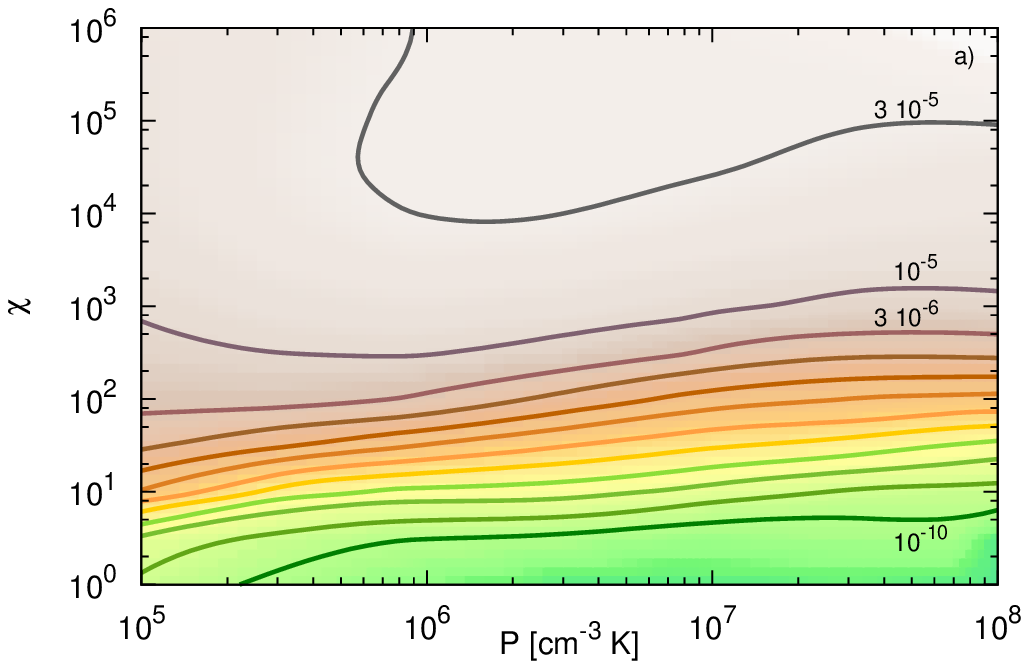}\includegraphics[bb=60bp 75bp 370bp 275bp,clip,width=0.5\linewidth]{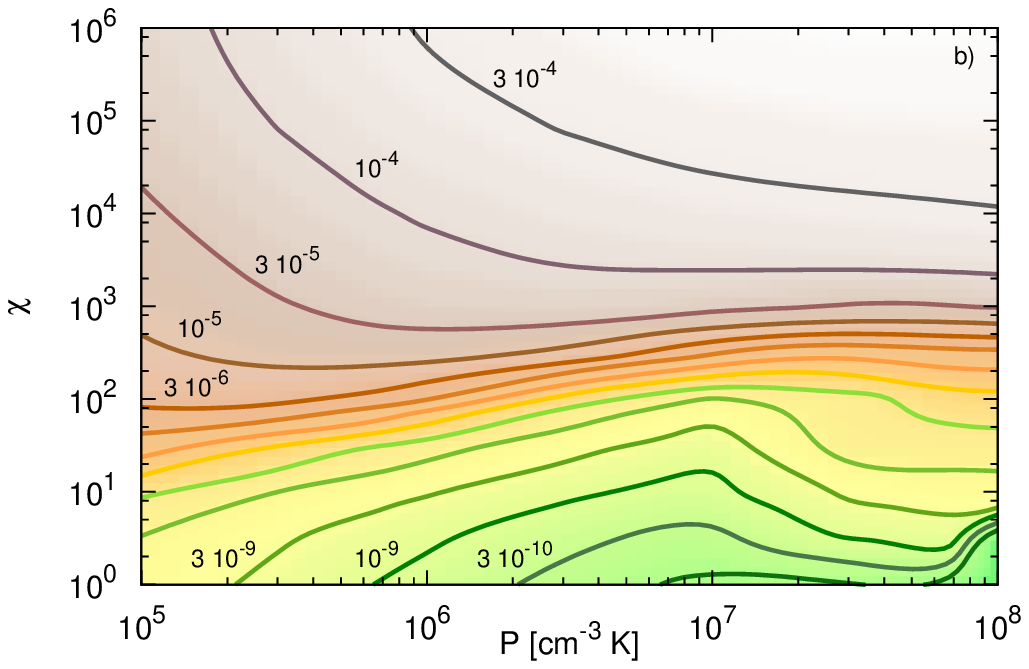}

\includegraphics[bb=60bp 75bp 370bp 275bp,clip,width=0.5\linewidth]{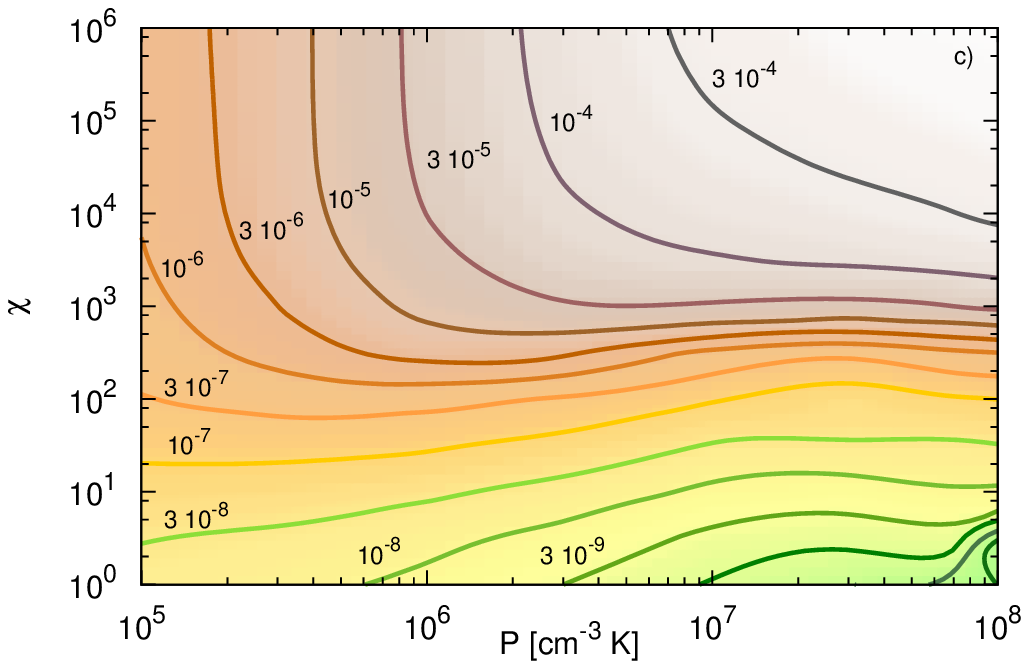}\includegraphics[bb=60bp 75bp 370bp 275bp,clip,width=0.5\linewidth]{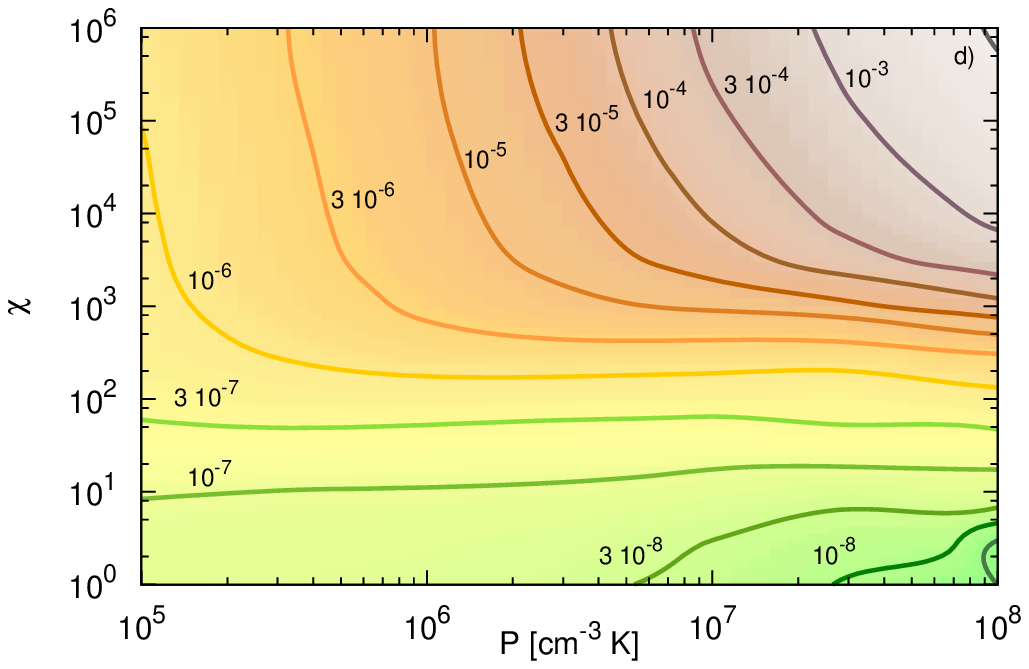}

\caption{Intensity of $\mathrm{H}_{2}$ lines in $\mathrm{erg}\,\mathrm{cm}^{-2}\,\mathrm{s}^{-1}\,\mathrm{sr}^{-1}$
seen in a face on geometry, for model C, a) 0-0 S(0), b) 0-0 S(1),
c) 0-0 S(2), d) 0-0 S(3).\label{fig:Int_H2_0-0}}
\end{figure*}

\begin{figure*}
\includegraphics[bb=60bp 75bp 370bp 275bp,clip,width=0.5\linewidth]{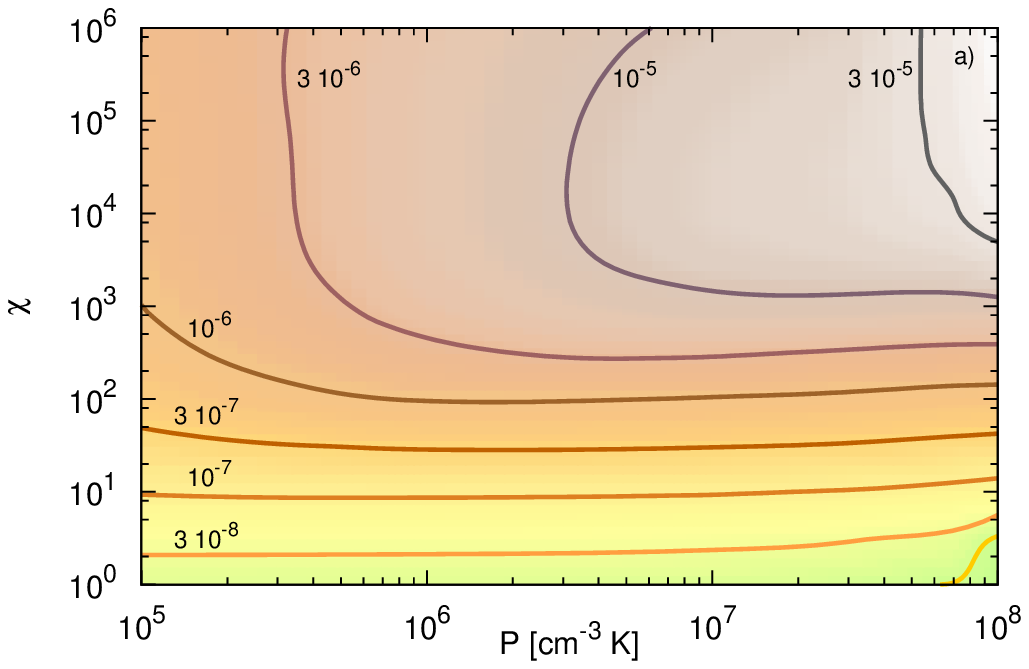}\includegraphics[bb=60bp 75bp 370bp 275bp,clip,width=0.5\linewidth]{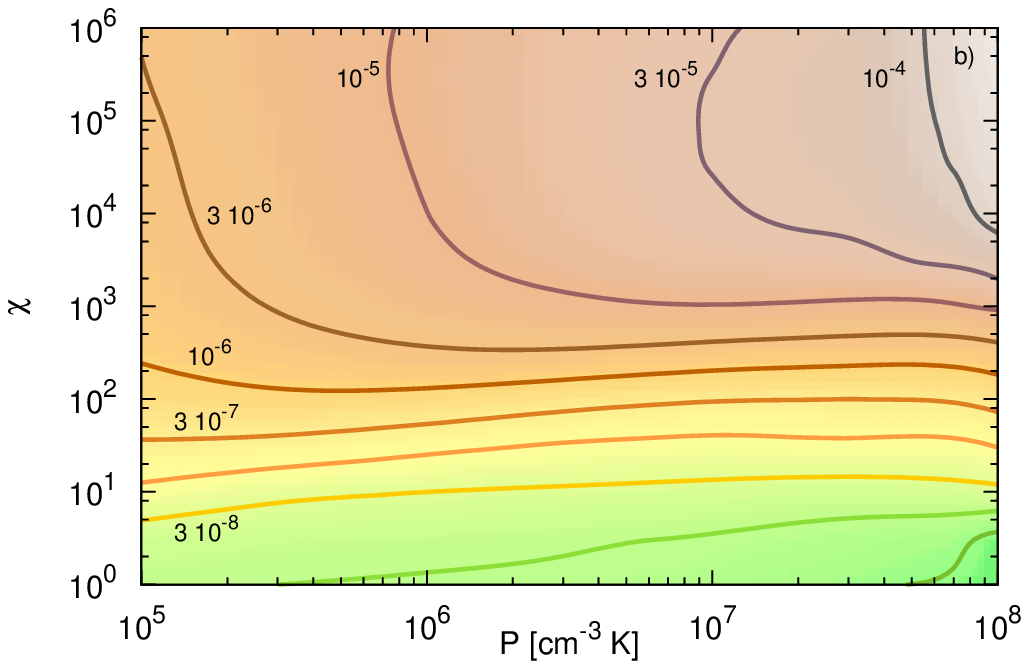}

\includegraphics[bb=60bp 75bp 370bp 275bp,clip,width=0.5\linewidth]{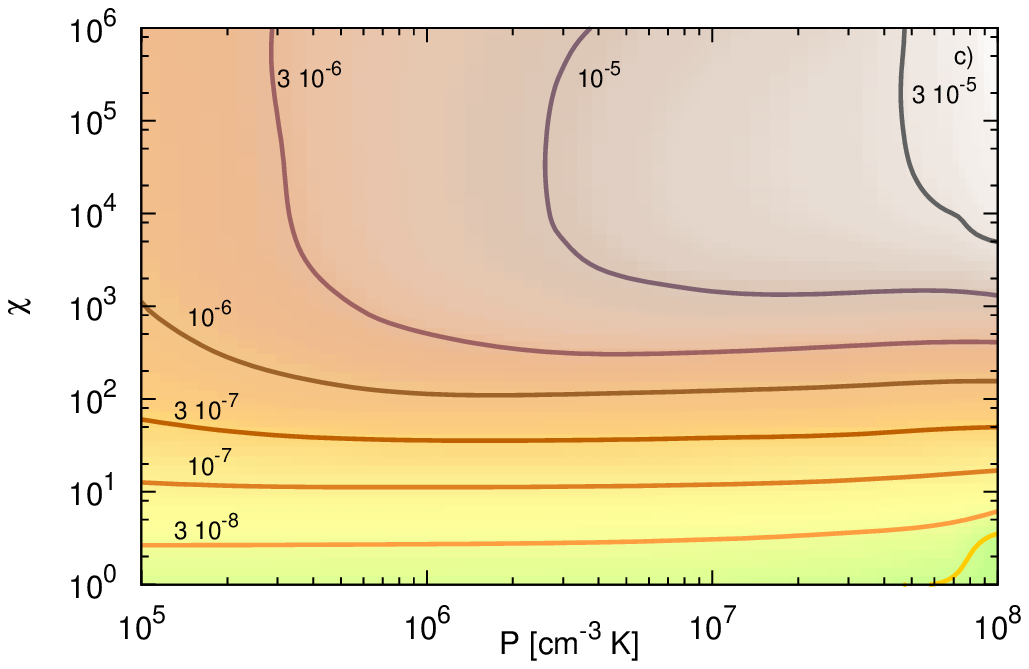}\includegraphics[bb=60bp 75bp 370bp 275bp,clip,width=0.5\linewidth]{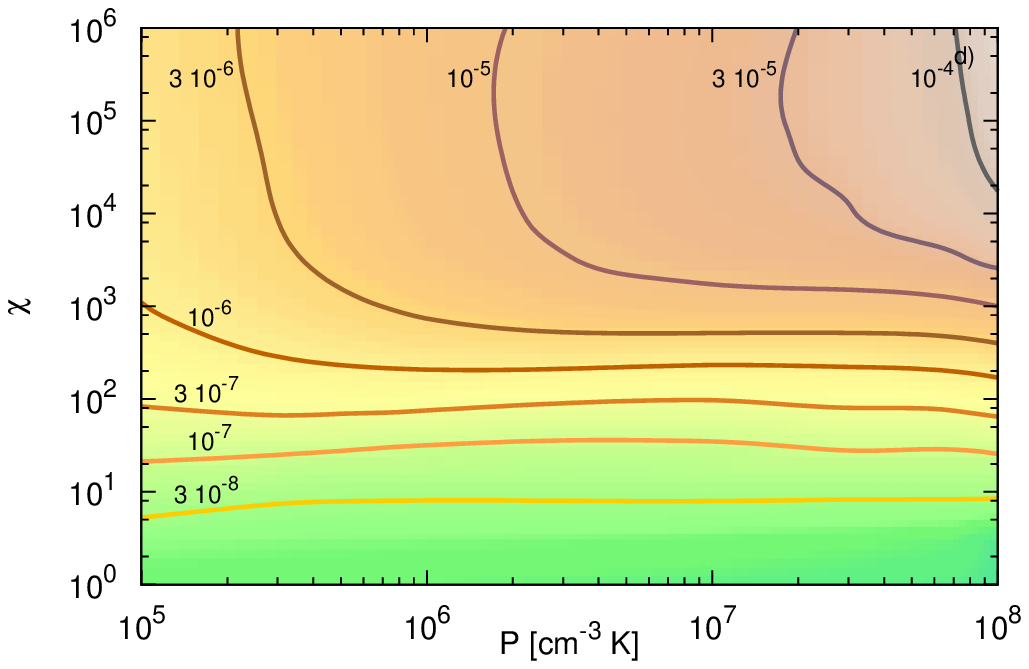}

\caption{Intensity of $\mathrm{H}_{2}$ lines in $\mathrm{erg}\,\mathrm{cm}^{-2}\,\mathrm{s}^{-1}\,\mathrm{sr}^{-1}$
seen in a face on geometry, for model C, a) 1-0 S(0), b) 1-0 S(1),
c) 1-0 S(2), d) 1-0 S(3).\label{fig:Int_H2_1-0}}
\end{figure*}

\begin{figure*}
\includegraphics[bb=60bp 75bp 370bp 275bp,clip,width=0.5\linewidth]{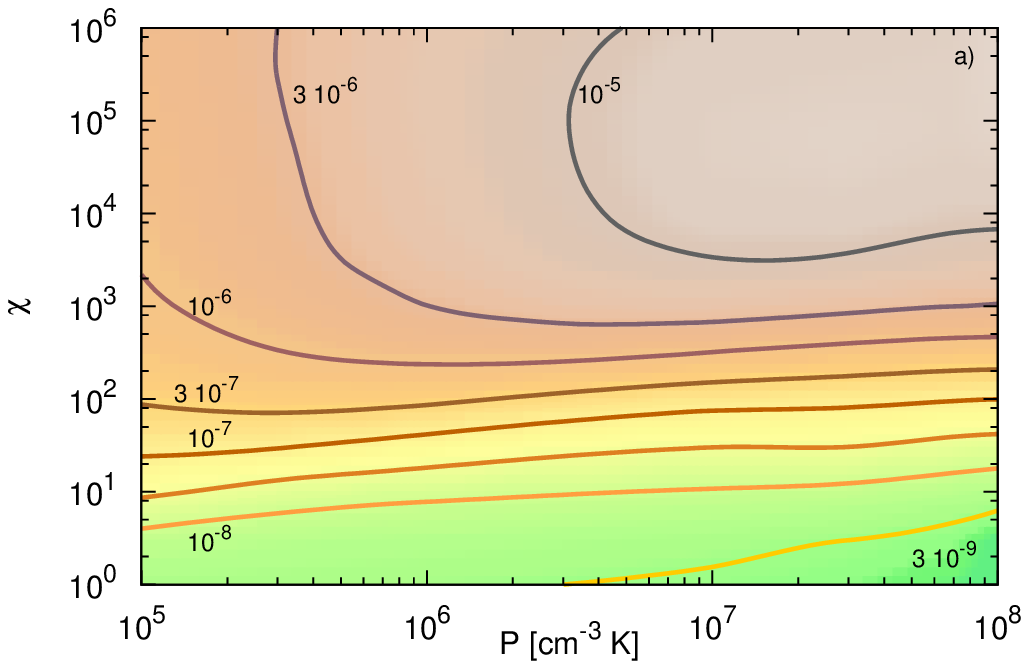}\includegraphics[bb=60bp 75bp 370bp 275bp,clip,width=0.5\linewidth]{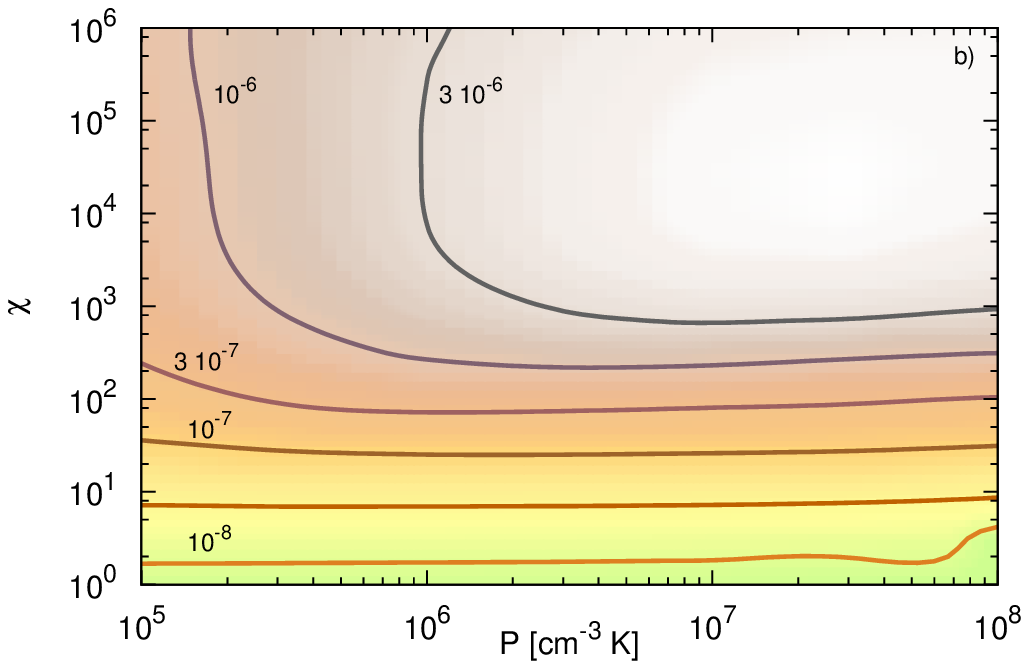}

\includegraphics[bb=60bp 75bp 370bp 275bp,clip,width=0.5\linewidth]{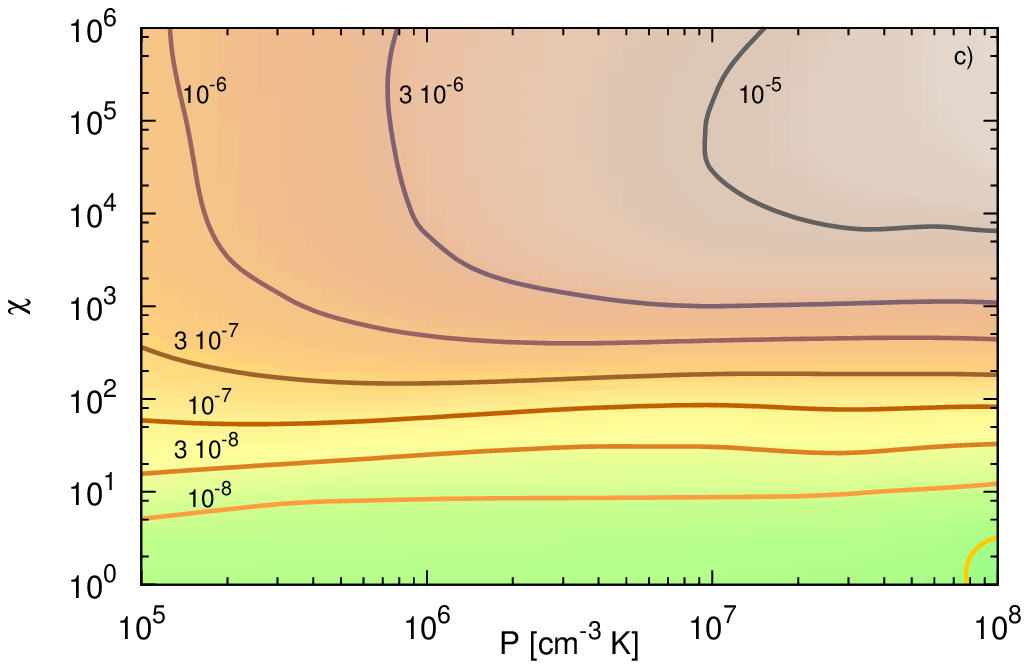}

\caption{Intensity of $\mathrm{H}_{2}$ lines in $\mathrm{erg}\,\mathrm{cm}^{-2}\,\mathrm{s}^{-1}\,\mathrm{sr}^{-1}$
seen in a face on geometry, for model C, a) 2-1 S(1), b) 2-1 S(2),
c) 2-1 S(3).\label{fig:Int_H2_2-1}}
\end{figure*}

We now investigate the implications of the new formation scenario
on the various emissivities. We display in Tables~\ref{tab:Emissivities}
and \ref{tab:Emissivities-H2} the emissivities derived from the previous
models. We focus on transitions emitted in the PDR region where photo-chemical
and -physical effects are taking place. In line with the results already
displayed, we find that the fine structure emission lines are not
very dependent on the formation model of $\mathrm{H_{2}}$ except
in a few cases, where temperatures become of the order of thousands
$\mathrm{K}$, due to the enhanced heating mechanisms described in
Sect.~\ref{sub:Heating}. Much more sensitive are the molecular hydrogen
emission transitions. We display the emissivities of the pure rotational
lines detected routinely by Spitzer as well as the 1-0 S(1) transition
at 2.12 $\mu$m detected from the ground. Ratios to other vibrational
lines detected in the K band are given to compare with past \citep{burton:02}
and future observations. We have also run 190 type C models with different
pressures ($10^{5}-10^{8}\,\mathrm{cm^{-3}\, K}$) and intensity scaling
factors of the incident radiation field ($1$ to $10^{6}$ Mathis
ISRF) on the observer side to display the corresponding contours in
different Figures. The back side radiation field is set to $\chi=1$
and the total visual magnitude, $\textrm{A}_{\textrm{V}}^{tot}=100$
in order to neglect radiative effects from the back side. The other
model parameters are given in Table~\ref{tab:ParametreModeles} and
the dust properties correspond to the standard values displayed in
Table~\ref{tab:Dust-properties.}. We display results when both LH
and ER mechanisms are activated for the formation of $\mathrm{H_{2}}$
(C type). Figs.~\ref{fig:Int_H2_0-0} to \ref{fig:intCp158} display
$\mathrm{H_{2}}$, $\mathrm{C^{+}}$ and $\mathrm{O}$ line intensities
contours in $\mathrm{erg}\,\mathrm{cm}^{-2}\,\mathrm{s}^{-1}\,\mathrm{sr}^{-1}$
for a face-on geometry. 

Intensity maps for other line intensities and other observation angles
can be found on our VO-Theory database, PDR Database (PDRDB), under
the project ``H2 formation LH+ER - 2011'' (\url{http://pdr.obspm.fr}).
Complete outputs of the models (giving access to abundances profiles,
temperature profile, formation and destruction rates, ... ), input
parameters and source code used to produce these models are also available
through PDRDB. 

Figs.~\ref{fig:Int_H2_0-0}, \ref{fig:Int_H2_1-0} and \ref{fig:Int_H2_2-1}
show that intensities (excepted for the 0-0 S(0) line) are mostly
dependent on $P$ for high values of $\chi/P$ and on $\chi$ for
low values of $\chi/P$. \citet{habart:04} explain this behavior
in the case of isochoric models. At the \textbf{$\mathrm{H}/\mathrm{H_{2}}$}
transition, for high $\chi/n_{H}$, photons are mostly absorbed by
grains whereas for low $\chi/n_{H}$, they are absorbed in \textbf{$\mathrm{H_{2}}$}
lines (self-shielding). It can be shown that in the first case, \textbf{$\mathrm{H_{2}}$}
line intensities are proportional to the\textbf{ $\mathrm{H_{2}}$}
formation rate and so to $n_{\mathrm{H}}$ 
 whereas, in the second case, they are proportional to the intensity
of the incident radiation field. In the case of isobaric models, an
analytical derivation is less obvious, but, as we can see on intensity
maps, this trend is valid over three orders of magnitudes of $P$.
In this $P$ range, $T$ varies only by a factor lower than 10 (Tables~\ref{tab:p5}
to \ref{tab:p8}). So, the variation of $P$ is essentially dominated
by the variation of $n_{\mathrm{H}}$, and the same argument applies.

The orders of magnitude of $\textrm{H}_{2}$ line intensities are
in good agreement with observations by Spitzer as presented in \citet{habart:11}.
For example, comparisons of our intensity maps to observations towards
the Horsehead for $\mathrm{H_{2}}$ lines \citep{habart:11} and $\mathrm{O}$
at $63\,\mu m$ \citep{Goicoechea09} gives $\chi$ of a few hundreds
in Mathis units which is consistent with previous models. Gas pressure
is more difficult to constrain this way since, for moderate $\chi$,
intensities of those lines are poorly dependent on $P$. Of all $\mathrm{H_{2}}$
lines 0-0 S(3) is the more difficult to reproduce. We also notice
that the intensity of the 1-0 S(1) line as well as the ratios of vibrational
transitions displayed in table~\ref{tab:Emissivities-H2} as potential
tests of formation pumping are within values reported\textbf{ }for
R1 and R2 towards M17 by \citet{burton:02} for a pressure of $10^{6}\,\mathrm{cm^{-3}}\,\mathrm{K}$
and radiation field enhancement factor of $10^{4}$. However the corresponding
modeled R3 ratio is significantly smaller. Models in this paper have
been done using typical properties of the interstellar medium (as
grains properties) that may not be applicable for a detailed interpretation
of the observations towards specific line of sights\textbf{.} As an
example, the star illuminating the Horsehead is a O9.5 type star with
an effective temperature of $33000\,\mathrm{K}$. Scaling a Mathis
ISRF as done here does not spread the UV energy in the same way than
a stellar spectrum. This can affect level excitation. Note that most
of the PDRs observations presented in \citet{habart:11} are seen
edge on whereas with our plane parallel model we provide in this paper
line intensities for PDRs seen face on and in PDRDBs up to 60 degrees.\textbf{
}Applications to specific lines of sight will be performed\textbf{
}subsequently by including all possible constraints derived from the
observations.

\begin{figure*}
\includegraphics[bb=60bp 75bp 370bp 275bp,clip,width=0.5\linewidth]{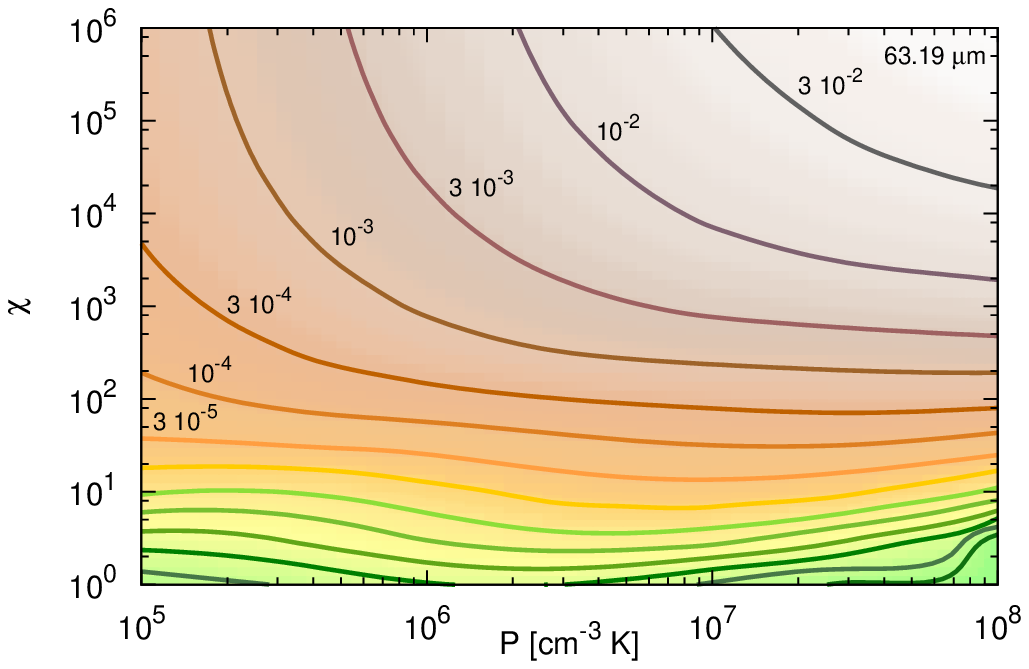}\includegraphics[bb=60bp 75bp 370bp 275bp,clip,width=0.5\linewidth]{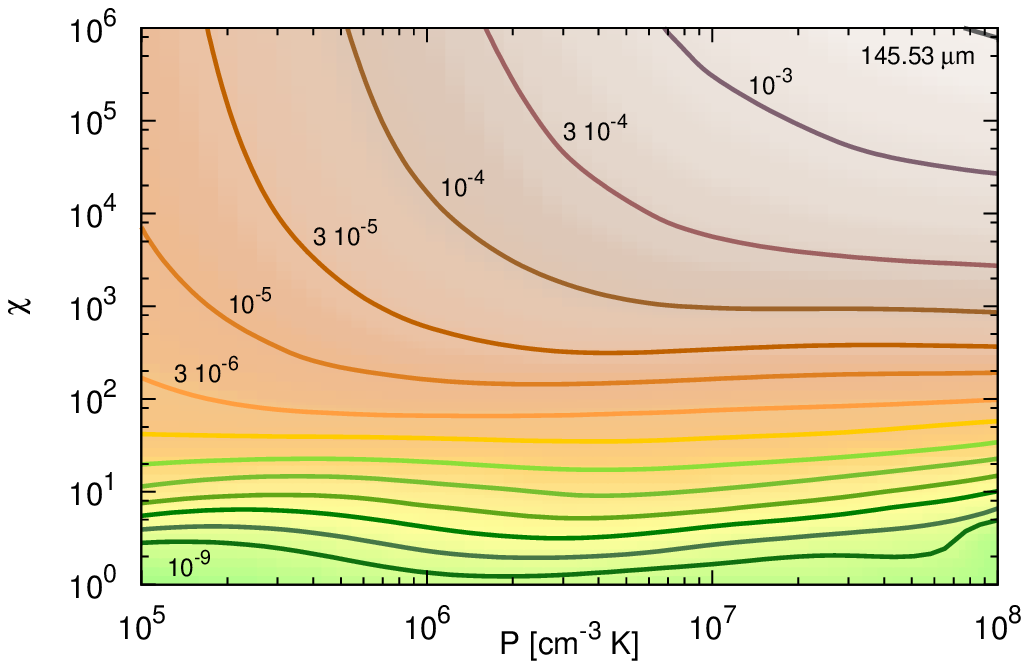}

\caption{Intensity of $\mathrm{O}$ at $63$ and $145\,\mu\textrm{m}$ in $\mathrm{erg}\,\mathrm{cm}^{-2}\,\mathrm{s}^{-1}\,\mathrm{sr}^{-1}$
seen in a face on geometry, for model C.}
\end{figure*}

\begin{figure*}
\includegraphics[bb=60bp 75bp 370bp 275bp,clip,width=0.5\linewidth]{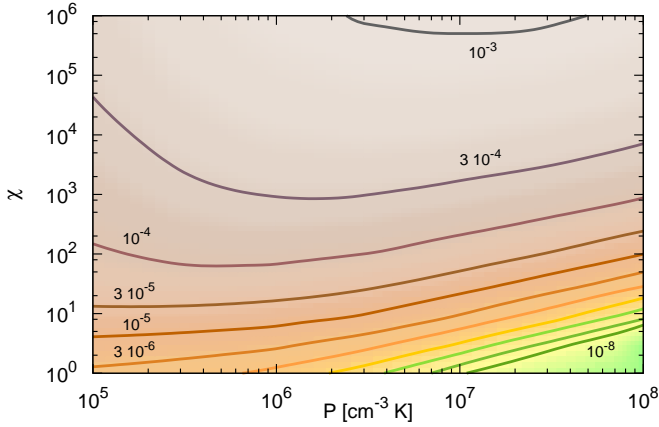}

\caption{Intensity of $\textrm{C}^{+}$ at $157.63\,\mu\textrm{m}$ in $\mathrm{erg}\,\mathrm{cm}^{-2}\,\mathrm{s}^{-1}\,\mathrm{sr}^{-1}$
seen in a face on geometry, for model C.\label{fig:intCp158}}
\end{figure*}

\begin{table*}
\caption{Emissivities in $\mathrm{erg}\,\mathrm{cm}^{-2}\,\mathrm{s}^{-1}\,\mathrm{sr}^{-1}$.
Numbers in parenthesis refer to powers of ten. \label{tab:Emissivities}}
\centering%
\begin{tabular}{>{\raggedright}p{1.5cm}>{\centering}p{2cm}ccccccccc}
\cline{2-11} 
\hline $p$ & $\chi_{obs}$ & \multicolumn{3}{c}{$10^{2}$} & \multicolumn{3}{c}{$10^{3}$} & \multicolumn{3}{c}{$10^{4}$}\tabularnewline
$(\mathrm{cm}^{-3}\,\mathrm{K})$ & Model & A & B & C & A & B & C & A & B & C\tabularnewline
\hline 
 & CII 158~$\mu\mathrm{m}$ & 9.1(-5) & 9.6(-5) & 8.6(-5) & 1.7(-4) & 1.9(-4) & 1.6(-4) & 2.5(-4) & 2.9(-4) & 2.4(-4)\tabularnewline
 & CI 610~$\mu\mathrm{m}$ & 2.4(-6) & 2.4(-6) & 2.4(-6) & 2.9(-6) & 3.1(-6) & 3.1(-6) & 3.5(-6) & 4.7(-6) & 4.8(-6)\tabularnewline
$10^{5}$ & CI 370~$\mu\mathrm{m}$ & 3.4(-6) & 3.4(-6) & 3.4(-6) & 4.7(-6) & 4.9(-6) & 4.9(-6) & 6.4(-6) & 7.0(-6) & 7.5(-6)\tabularnewline
 & OI 145~$\mu\mathrm{m}$ & 2.2(-6) & 2.0(-6) & 2.2(-6) & 6.3(-6) & 5.7(-6) & 6.3(-6) & 1.0(-5) & 9.3(-5) & 1.0(-4)\tabularnewline
 & OI 63~$\mu\mathrm{m}$ & 6.4(-5) & 6.4(-5) & 6.1(-5) & 1.8(-4) & 1.8(-4) & 1.7(-4) & 2.9(-4) & 3.2(-4) & 2.9(-4)\tabularnewline
\hline 
 & CII 158~$\mu\mathrm{m}$ & 1.3(-4) & 1.3(-4) & 1.3(-4) & 3.2(-4) & 3.2(-4) & 3.1(-4) & 5.2(-4) & 5.4(-4) & 5.1(-4)\tabularnewline
 & CI 610~$\mu\mathrm{m}$ & 2.7(-6) & 2.7(-6) & 2.7(-6) & 3.1(-6) & 3.4(-6) & 3.4(-6) & 3.7(-6) & 4.3(-6) & 4.4(-6)\tabularnewline
$10^{6}$ & CI 370~$\mu\mathrm{m}$ & 5.7(-6) & 5.6(-6) & 5.6(-6) & 7.6(-6) & 8.3(-6) & 8.3(-6) & 1.0(-5) & 1.1(-5) & 1.2(-5)\tabularnewline
 & OI 145~$\mu\mathrm{m}$ & 5.7(-6) & 4.1(-6) & 5.8(-6) & 3.8(-5) & 3.2(-5) & 3.9(-5) & 8.4(-5) & 7.2(-5) & 8.7(-5)\tabularnewline
 & OI 63~$\mu\mathrm{m}$ & 1.9(-4) & 1.9(-4) & 1.8(-4) & 1.1(-3) & 1.1(-3) & 9.8(-4) & 2.3(-3) & 2.4(-3) & 2.2(-3)\tabularnewline
\hline 
 & CII 158~$\mu\mathrm{m}$ & 5.8(-5) & 5.7(-5) & 5.7(-5) & 2.4(-4) & 2.3(-4) & 2.4(-4) & 2.4(-4) & 4.9(-4) & 5.1(-4)\tabularnewline
 & CI 610~$\mu\mathrm{m}$ & 2.8(-6) & 2.8(-6) & 2.8(-6) & 3.3(-6) & 3.5(-6) & 3.5(-6) & 3.3(-6) & 3.8(-6) & 4.2(-6)\tabularnewline
$10^{7}$ & CI 370~$\mu\mathrm{m}$ & 6.7(-6) & 6.7(-6) & 6.7(-6) & 9.6(-6) & 1.0(-5) & 1.0(-5) & 9.6(-6) & 1.3(-5) & 1.5(-5)\tabularnewline
 & OI 145~$\mu\mathrm{m}$ & 4.9(-6) & 3.8(-6) & 4.3(-6) & 9.0(-5) & 5.7(-5) & 9.9(-5) & 9.0(-5) & 2.7(-4) & 3.6(-4)\tabularnewline
 & OI 63~$\mu\mathrm{m}$ & 3.6(-4) & 3.5(-4) & 3.5(-4) & 3.4(-3) & 3.3(-3) & 3.3(-3) & 3.4(-3) & 1.1(-2) & 1.0(-2)\tabularnewline
\hline 
 & CII 158~$\mu\mathrm{m}$ & 6.5(-6) & 1.0(-5) & 9.8(-6) & 1.1(-4) & 1.2(-4) & 1.1(-4) & 3.5(-4) & 3.8(-4) & 3.3(-4)\tabularnewline
 & CI 610~$\mu\mathrm{m}$ & 2.2(-6) & 2.3(-6) & 2.3(-6) & 3.0(-6) & 3.3(-6) & 3.3(-6) & 3.4(-6) & 3.6(-6) & 4.2(-6)\tabularnewline
$10^{8}$ & CI 370~$\mu\mathrm{m}$ & 5.8(-6) & 5.7(-6) & 5.7(-6) & 1.0(-5) & 1.1(-5) & 1.2(-5) & 1.4(-5) & 1.5(-5) & 1.8(-5)\tabularnewline
 & OI 145~$\mu\mathrm{m}$ & 3.5(-5) & 2.4(-6) & 3.0(-6) & 9.9(-5) & 7.3(-5) & 1.1(-4) & 5.7(-4) & 4.4(-4) & 6.1(-4)\tabularnewline
 & OI 63~$\mu\mathrm{m}$ & 1.7(-3) & 3.7(-4) & 3.9(-4) & 5.7(-3) & 4.9(-3) & 5.9(-3) & 2.3(-2) & 2.1(-2) & 2.3(-2)\tabularnewline
\hline 
\end{tabular}
\end{table*}

\begin{table*}
\caption{Emissivities of\textbf{ $\mathrm{H_{2}}$ }transitions in $\mathrm{erg}\,\mathrm{cm}^{-2}\,\mathrm{s}^{-1}\,\mathrm{sr}^{-1}$.
Number in parenthesis refer to powers of ten.\protect \\
$R_{1}:$ $1-0S(1)/2-1S(1)$, $R_{2}:$ $1-0S(1)/1-0S(7)$, $R_{3}:$
$1-0S(1)/6-4O(3)$,\label{tab:Emissivities-H2}}
\centering%
\begin{tabular}{>{\raggedright}p{1.1cm}>{\centering}p{2cm}ccccccccc}
\hline 
\hline $p$ & $\chi_{obs}$ & \multicolumn{3}{c}{$10^{2}$} & \multicolumn{3}{c}{$10^{3}$} & \multicolumn{3}{c}{$10^{4}$}\tabularnewline
$(\mathrm{cm}^{-3}\,\mathrm{K})$ & Model & A & B & C & A & B & C & A & B & C\tabularnewline
\hline 
 & 0-0 S(0) & 2.1(-6) & 4.4(-7) & 3.7(-6) & 7.0(-6) & 2.3(-7) & 1.1(-5) & 1.4(-5) & 2.1(-10) & 1.7(-5)\tabularnewline
 & 0-0 S(1) & 1.2(-6) & 1.2(-7) & 3.5(-6) & 5.8(-6) & 4.0(-8) & 1.2(-5) & 1.5(-5) & 6.0(-10) & 2.4(-5)\tabularnewline
 & 0-0 S(2) & 1.5(-7) & 7.1(-8) & 2.6(-7) & 2.8(-7) & 3.2(-8) & 6.4(-7) & 5.1(-7) & 9.8(-10) & 9.8(-7)\tabularnewline
 & 0-0 S(3) & 2.1(-7) & 0.5(-8) & 3.4(-7) & 3.2(-7) & 4.0(-8) & 6.7(-7) & 4.0(-7) & 4.4(-10) & 8.0(-7)\tabularnewline
$10^{5}$ & 1-0 S(1) & 4.0(-7) & 1.6(-7) & 6.2(-7) & 8.1(-7) & 5.9(-8) & 1.6(-6) & 1.1(-6) & 1.2(-9) & 2.1(-6)\tabularnewline
 & $R_{1}$ & 2.0 & 2.0 & 2.0 & 1.9 & 2.0 & 1.9 & 1.9 & 1.9 & 1.9\tabularnewline
 & $R_{2}$ & 7.5 & 7.1 & 5.4 & 7.7 & 8.1 & 4.9 & 7.4 & 93 & 4.8\tabularnewline
 & $R_{3}$ & 3.6 & 3.6 & 3.5 & 4.0 & 3.8 & 3.9 & 4.0 & 4.2 & 4.0\tabularnewline
\hline 
 & 0-0 S(0) & 1.9(-6) & 3.3(-7) & 2.3(-6) & 1.2(-5) & 3.2(-7) & 1.8(-5) & 2.2(-5) & 1.3(-9) & 2.9(-5)\tabularnewline
 & 0-0 S(1) & 6.2(-7) & 6.4(-7) & 9.7(-7) & 1.4(-5) & 3.2(-8) & 4.4(-5) & 4.7(-5) & 1.2(-9) & 1.1(-4)\tabularnewline
 & 0-0 S(2) & 3.8(-7) & 1.9(-7) & 4.5(-7) & 3.1(-5) & 9.4(-8) & 1.3(-5) & 1.1(-5) & 4.7(-9) & 2.8(-5)\tabularnewline
 & 0-0 S(3) & 4.7(-7) & 2.5(-7) & 5.3(-7) & 1.4(-6) & 1.1(-7) & 3.4(-6) & 2.1(-6) & 2.9(-9) & 5.6(-6)\tabularnewline
$10^{6}$ & 1-0 S(1) & 6.7(-7) & 3.2(-7) & 7.4(-7) & 2.8(-6) & 1.5(-7) & 5.6(-6) & 4.5(-6) & 6.5(-9) & 9.2(-6)\tabularnewline
 & $R_{1}$ & 2.0 & 2.0 & 2.0 & 2.0 & 2.0 & 1.9 & 1.9 & 2.0 & 1.9\tabularnewline
 & $R_{2}$ & 4.2 & 3.9 & 3.2 & 5.2 & 4.9 & 3.7 & 5.2 & 22 & 3.4\tabularnewline
 & $R_{3}$ & 4.1 & 4.8 & 4.0 & 4.2 & 5.0 & 4.0 & 4.2 & 6.3 & 4.1\tabularnewline
\hline 
 & 0-0 S(0) & 2.2(-7) & 1.1(-7) & 1.7(-7) & 7.9(-6) & 2.8(-7) & 1.1(-5) & 1.9(-5) & 4.7(-9) & 2.4(-5)\tabularnewline
 & 0-0 S(1) & 1.2(-8) & 7.8(-9) & 9.5(-9) & 1.0(-5) & 2.7(-8) & 3.6(-5) & 7.3(-5) & 1.0(-9) & 1.9(-4)\tabularnewline
 & 0-0 S(2) & 1.1(-7) & 1.2(-7) & 1.1(-7) & 5.4(-6) & 2.5(-7) & 2.4(-5) & 4.5(-5) & 1.5(-8) & 1.4(-4)\tabularnewline
 & 0-0 S(3) & 4.8(-7) & 3.9(-7) & 4.5(-7) & 4.5(-6) & 5.6(-7) & 1.1(-5) & 1.4(-5) & 1.9(-8) & 8.5(-5)\tabularnewline
$10^{7}$ & 1-0 S(1) & 3.7(-7) & 2.9(7) & 3.4(-7) & 5.8(-6) & 8.7(-7) & 9.6(-6) & 1.3(-5) & 2.9(-8) & 2.5(-5)\tabularnewline
 & $R_{1}$ & 2.2 & 2.3 & 2.1 & 2.1 & 2.2 & 2.1 & 2.1 & 2.3 & 2.1\tabularnewline
 & $R_{2}$ & 1.4 & 1.4 & 1.2 & 2.6 & 2.2 & 2.2 & 3.0 & 6.5 & 2.2\tabularnewline
 & $R_{3}$ & 7.0 & 11 & 8.2 & 5.2 & 9.3 & 4.4 & 4.7 & 15 & 4.4\tabularnewline
\hline 
 & 0-0 S(0) & 3.8(-6) & 4.7(-8) & 6.9(-8) & 6.2(-6) & 1.6(-7) & 7.2(-6) & 1.6(-5) & 1.8(-8) & 1.8(-5)\tabularnewline
 & 0-0 S(1) & 2.3(-6) & 1.1(-9) & 7.4(-8) & 1.1(-5) & 2.5(-9) & 2.9(-5) & 1.4(-4) & 6.2(-10) & 2.6(-4)\tabularnewline
 & 0-0 S(2) & 1.4(-5) & 1.5(-8) & 1.1(-7) & 1.1(-5) & 4.0(-8) & 3.2(-5) & 1.3(-4) & 1.3(-8) & 2.9(-4)\tabularnewline
 & 0-0 S(3) & 1.3(-5) & 2.0(-7) & 7.6(-7) & 7.7(-6) & 2.8(-7) & 5.7(-5) & 2.8(-4) & 4.1(-8) & 1.0(-3)\tabularnewline
$10^{8}$ & 1-0 S(1) & 1.3(-5) & 2.0(-7) & 4.7(-7) & 5.3(-6) & 3.4(-7) & 1.1(-5) & 2.2(-5) & 8.2(-8) & 9.7(-5)\tabularnewline
 & $R_{1}$ & 2.9 & 2.3 & 4.7 & 2.8 & 2.7 & 4.0 & 4.0 & 3.3 & 9.8\tabularnewline
 & $R_{2}$ & 0.4 & 2.0 & 0.9 & 0.98 & 0.9 & 1.2 & 2.0 & 3.3 & 3.4\tabularnewline
 & $R_{3}$ & 33 & 18 & 18 & 8.6 & 27 & 8.7 & 8.1 & 44 & 20\tabularnewline
\hline 
\end{tabular}
\end{table*}

\section{Conclusions\label{sec:Conclusions}}

In this paper, we presented implementations of Langmuir-Hinshelwood
and Eley-Rideal mechanisms to describe\textbf{ $\mathrm{H_{2}}$ }formation\textbf{
}on grain surfaces within\textbf{ }the Meudon PDR code. This is the
first of a series intended to discuss the formation of other molecules
on grains\textbf{ }in the light of recent experimental and theoretical
progresses. Eley-Rideal mechanism involving chemisorbed sites\textbf{
}is an efficient process to form $\mathrm{H}_{2}$ in regions where
the gas is hot and dust grains are warm\textbf{. }In previous versions
of the code, the sole Langmuir-Hinshelwood mechanism allows to recover
the value of the mean formation rate of $\mathrm{H}_{2}$ derived
from VUV observations of diffuse and translucent clouds. However,
it led to strongly reduced values of $\mathrm{H}_{2}$ formation rates
in denser and strongly illuminated clouds due to the increasing importance
of $\mathrm{H}$ evaporation compared to diffusion at the surface
of the dust grains. Observations of these high excitation PDRs by,
e.g., \citet{habart:11}\textbf{ }and references therein suggest mean
formation rates typically a factor of 2 to 5 higher (with large uncertainties).
This is now recovered by our treatment of the Eley-Rideal process
whose efficiency rises with increasing gas temperature (up to a threshold).
Thus, considering that observed rates are mean values along the line
of sight and across a large variety of physical conditions wether
our theoretical rates are local values computed from micro-physics
considerations alone, the agreement between model and observations
is now quite satisfactory in high excitation regions too.

The Eley-Rideal mechanism also provides an efficient process to heat
the gas. Following \citet{sizun:10} prescription to spread the energy
released by the formation of a $\mathrm{H}_{2}$ molecule in kinetic
energy and internal energy, we showed that this formation process
can be the major heating mechanism under some specific conditions.
In particular, in regions where $\mathrm{H}_{2}$ self-shielding is
not yet fully efficient and with densities above the critical density
of collisions with $\mathrm{H}$, a significant amount of energy can
be transferred into kinetic energy of the gas.\textbf{ }Signatures
of formation pumping are not straightforward at the light of present
knowledge on the internal state of newly formed $\mathrm{H}_{2}$.

The efficiency of the Langmuir-Hinshelwood process depends on the
temperature of the grains. As shown by many authors this formation
mechanism can only take place in a small window of grain temperatures
\citep{biham:02,biham:05,lepetit:09}, 10-20 K for amorphous carbons.
As a consequence, LH mechanism becomes the major\textbf{ $\mathrm{H}_{2}$}
formation process at moderate visual extinctions. The strong dependence
on $T_{gr}$ implies that a proper computing of grains temperature
has to be done in numerical models, including temperature fluctuation
of the smallest grains.

We compared results of PDR models in which $\mathrm{H}_{2}$ formation
is treated by these detailed mechanisms to models using a constant
formation rate with $R_{\mathrm{H_{2}}}=3\times10^{-17}\mathrm{cm}^{3}\,\mathrm{s}^{-1}$.
For diffuse clouds parameters, the main effect is to shift the $\mathrm{H}/\mathrm{H}_{2}$
transition and to increase the molecular fraction. For stronger PDRs,
we showed that at high radiation field, line intensities depend strongly
on gas pressure. The difference between detailed treatment of $\mathrm{H}_{2}$
formation and a constant formation rate can lead to differences up
to a factor 3 in the most commonly observed $\mathrm{H}_{2}$ line
intensities. We provide maps of such line intensities for a set of
190 models for different pressures and intensities of the incident
radiation field%
\footnote{Full results are found at \url{http://pdr.obspm.fr}%
}.
\begin{acknowledgements}
This work was \textbf{partially} funded by grant ANR-09-BLAN-0231-01
from the French Agence Nationale de la Recherche as part of the SCHISM
project\textbf{ }and\textbf{ }by the french CNRS national program
PCMI\textbf{.} The authors acknowledge support for computing resources
and services from France Grilles and the EGI e-infrastructure as well
as MESOPSL, the computing center for Paris Sciences et Lettres. Some
kinetic data have been downloaded from the online KIDA (KInetic Database
for Astrochemistry,\url{ http://kida.obs.u-bordeaux1.fr}) database.
We thank our colleagues from the ``Laboratoire des Collisions Atomiques
et Moléculaires'' in Orsay, V. Sidis, N. Rougeaud and D. Bachellerie
for fruitful discussions on the ER process.\textbf{ }We thank David
Languignon, Nicolas Moreau, Benjamin Ooghe and Emeric Bron for their
help in running some of the models and setting them up on the on-line
database.
\end{acknowledgements}
\appendix

\section{Other recent updates to the PDR code\label{sec:Other-recent-updates}}

Different progresses have been achieved in computing fine structure
and rotational excitation due to collisions with $\mathrm{H}$ and/or
$\mathrm{H_{2}}$ as reported in the BASECOL\citep{dubernet:06} and
LAMDA \citep{schoier:05} databases. We have updated and/or implemented
collisional excitation rates which play an active role in the cooling
processes and compute explicitly their emission spectrum by solving
the statistical equilibrium equations, including radiative pumping
by the cosmic background radiation field and dust infrared emission
\citep{gonzalez:08}. Let us point out that the new fine structure
excitation collision rates of atomic oxygen computed by \citet{abraham:07}
have a significant impact on the temperature at the edge of PDRs.
As an example, for a typical proton density of $10^{4}\,\mathrm{cm}^{-3}$
and a radiation scaling factor of 10, the temperature at the edge
is 90K with the old \citet{launay:77} atomic oxygen collision rates
and only 67K with the recent values displayed in \citet{abraham:07}.
In the present stage, emission spectra of millimeter and submillimeter
transitions of $\mathrm{HCN}$, $\mathrm{OH}$, $\mathrm{CH^{+}}$,
$\mathrm{O_{2}}$ are readily computed and implementation of other
molecules is straightforward and depends $ $\emph{only} on the availability
of collision rates by the relevant perturbers. Another significant
issue is the inclusion of the thermal and charge balance of the grains
in the overall ionization fraction. The thermal balance is obtained
through a coupling with the DustEM program \citet{compiegne:11} which
can be switched on through the F\_Dustem parameter (\texttt{F\_Dustem=1}).
If not (\texttt{F\_Dustem=0}), the temperature of the various grain
bin sizes is obtained from the formula (5) of \citet{hollenbach:91},
where the actual value of the radiation field is introduced. The determination
of the grain charge is obtained from the balance between photoelectric
effect and recombination on dust particles, expanding on the treatment
of \citet{draine:87} and \citet{bakes:94}.

We also compute photodissociation rates from the integration of the
photodissociation cross-sections, when available, with the interstellar
radiation field. The attenuation by dust particles is then directly
obtained from the dust properties considered in the model, i.e. their
absorption and extinction coefficients which depend on the size and
the nature of the dust particles. Different options are proposed depending
on the treatment of the grain temperatures. If \texttt{F\_Dustem=1},
we use the absorption and scattering coefficients computed by the
DustEM code \citep{compiegne:11}. If \texttt{F\_Dustem=0}, we derive
the albedo and dust properties from the extinction curve given by
the Fitzpatrick and Massa analytic expansion \citep{fitzpatrick:07},
extended towards longer wavelengths by the data from \citet{weingartner:01}
\footnote{available from \url{http://www.astro.princeton.edu/~draine/dust/dustmix.html}%
}. It is remarkable that we recover the dust free photodissociation
rates displayed in \citet{vd:88} for the Mathis or Draine incident
radiation field. The $A_{V}$ dependence of the photodissociation
rates reflects then directly the appropriate dust environment.

\section{\label{App_Langmuir}Langmuir-Hinshelwood mechanism}

Upon landing on a grain most heavy species may build an ice mantle.
In that case, it is possible to account for the total number of physisorbed
molecules by integrating on the grain size distribution. This is the
subject of a forthcoming paper on surface chemistry (Le Petit et al.,
to be submitted). However this is most probably not the case for the
lightest species ($\mathrm{H}$, $\mathrm{H}_{2}$, $\mathrm{D}$,
$\mathrm{HD}$, ...). We assume here that they only build a single
monolayer above either the grain surface or the ice mantle.

In that case, two effects must be taken into account:
\begin{itemize}
\item Upon landing on a site already occupied by a light species, the impinging
species is rejected to the gas phase.
\item Binding depends on the (size dependent) temperature of the grain and
thus steady state depends on the size (and characteristics) of the
grain.
\end{itemize}
So, we must compute the number of physisorbed particles of type $\mathrm{X}$
on a grain of size $a$. This is $N_{\mathrm{X:}}(a)$ in the following
(in particles per grain, and not in particles per cubic centimeter).
The total amount of $\mathrm{X}$ on all grains follows by integration:
\[
[X:]=\int_{a_{min}}^{a_{max}}\, N_{\mathrm{X:}}(a)\, dn_{g}=A_{gr}\, n_{\mathrm{H}}\,\int_{a_{min}}^{a_{max}}\, N_{\mathrm{X:}}(a)\, a^{-\alpha}\, da
\]
where the second expression is for a MRN size distribution with $dn_{g}=A_{gr}\, n_{\mathrm{H}}\, a^{-\alpha}\, da$;
where $A_{gr}$ is a normalization factor, $n_{\mathrm{H}}$ is the
gas density (in $\mathrm{cm}^{-3}$) and $a$ the grain radius (in
$\mathrm{cm}$). In the following, we will use that case as an example,
but it is easy to generalize to any distribution. Numerical integration
is performed by discretizing the size. When needed, we will use:
\[
\int_{a_{min}}^{a_{max}}\, f(a)\, dn_{g}=\sum_{i=1}^{npg}w_{i}\, f(a_{i})
\]
where the weights $w_{i}$ and abscissae $a_{i}$ are chosen according
to the distribution. The number of abscissae is $npg$. Table~\ref{tab_Gauss_Int}
displays the abscissae and weights computed for the parameters of
the MRN distribution given in \ref{tab:Dust-properties.}. These coefficients
must be computed anew if one changes the range of sizes%
\footnote{They are computed automatically during the initialization phase of
the code.%
}. Here (note that the power $a^{-3.5}$ does not appear in the discrete
sum):
\[
\int_{a_{min}}^{a_{max}}\, f(a)\, a^{-3.5}\, da=\sum_{i=1}^{12}\, w_{i}\, f(a_{i})
\]

\begin{table}
\caption{Abscissae and weights for the Gaussian integration of a MRN size distribution
with $npg=12$, $\alpha=3.5$, $a_{min}=3\,10^{-7}\,\mathrm{cm}$
and $a_{max}=3\,10^{-5}\,\mathrm{cm}$. The weights do not include
the normalization factor $A_{gr}$ nor the gas phase density $n_{\mathrm{H}}$.
In the text, $w_{i}=A_{gr}\, n_{\mathrm{H}}\, w'_{i}$.\label{tab_Gauss_Int}}

\centering%
\begin{tabular}{cc}
\hline 
\hline$a_{i}$ & $w'_{i}$\tabularnewline
\hline 
3.57699069369688527E-007  & 5792151389936155.0 \tabularnewline
6.98799352454772693E-007 & 2051685967306352.5\tabularnewline
1.66467328055428966E-006 & 235554962470170.16\tabularnewline
3.55645670102603961E-006 & 28071872396201.828\tabularnewline
6.38977929313982604E-006 & 4932081459260.9521\tabularnewline
9.97829390939405377E-006 & 1234971510565.4812\tabularnewline
1.40343774171740631E-005 & 403598849340.45422\tabularnewline
1.82200037492632748E-005 & 160061512934.20764\tabularnewline
2.21826037436107715E-005 & 72679832471.918198\tabularnewline
2.55871898379409544E-005 & 35689784319.592133\tabularnewline
2.81455972293716657E-005 & 17382741166.856663\tabularnewline
2.96417164382822754E-005 & 6457454266.3027544\tabularnewline
\hline 
\end{tabular}
\end{table}

Three types of reactions are to consider:
\begin{itemize}
\item Adsorption
\item Ejection
\item Reaction
\end{itemize}
We will not consider here reactions with heavy atoms or molecules
(including ices) that will be the subject of a follow-up paper.

\subsection{Adsorption}

If the outer layer of the grain is populated by light species ($\mathrm{H}$,
$\mathrm{H}_{2}$) then any of them may lead to rejection of an impinging
atom. Let us consider $n_{j}$ such species. Then, the number of accretion
of a species $\mathrm{X}$ per unit time interval on a single grain
of size $a$ is:
\[
s(\mathrm{X})\,[\mathrm{X}]\,\bar{v}(\mathrm{X})\,\pi\, a^{2}\,\left(1-\frac{d_{s}^{2}}{4\pi\, a^{2}}\,\sum_{j}^{n_{j}}N_{\mathrm{Y_{j}:}}(a)\right)
\]
where $s(\mathrm{X})$ is the sticking coefficient of species $\mathrm{X}$,
$d_{s}$ the mean distance between adsorption sites (supposed identical
for all grains) and the term in parentheses takes into account rejection
by any species $Y_{j}$ that is already on the grain. $\frac{4\pi\, a^{2}}{d_{s}^{2}}$
is the total number of adsorption sites on a grain of size $a$. Formally,
this equation may be split into a first order formation reaction with
rate $k_{ad}(a)=s(\mathrm{X})\,\bar{v}(\mathrm{X})\,\pi\, a^{2}$,
and $n_{j}$ different second order destruction reactions with rate
$k_{rej}=s(\mathrm{X})\,\bar{v}(\mathrm{X})\,\frac{d_{s}^{2}}{4}$.
The relevant creation and destruction equations are thus:
\begin{equation}
\frac{dN_{\mathrm{X:}}(a)}{dt}=k_{ad}(a)\,[\mathrm{X}]-\sum_{j}\, k_{rej}\,[\mathrm{X}]\, N_{\mathrm{Y_{j}:}}(a)\label{eq_surf_ads}
\end{equation}
\[
-\frac{d[\mathrm{X}]}{dt}=\left(\int_{a_{min}}^{a_{max}}\, k_{ad}\, dn_{g}\right)\,[\mathrm{X}]-\sum_{j}\,\int_{a_{min}}^{a_{max}}\, k_{rej}\, N_{\mathrm{Y_{j}:}}(a)\, dn_{g}\,[\mathrm{X}]
\]
or:
\begin{equation}
-\frac{d[\mathrm{X}]}{dt}=k_{rej}\,\frac{S_{gr}}{d_{s}^{2}}\,[\mathrm{X}]-k_{rej}\,\sum_{j}^{n_{j}}\,\sum_{i}^{npg}\, w_{i}\, N_{\mathrm{Y_{j}:}}(a_{i})\,[\mathrm{X}]\label{eq_adsor}
\end{equation}

Here, $S_{gr}$ is the total surface of grains per unit volume. One
can see that, although the form of the reaction terms is preserved
(first or second order polynomial in the variables) the total number
of individual contributions becomes large (one accretion leads to
$(n_{j}\times npg+1)$ reactions)%
\footnote{In the Meudon PDR code, we use a specific Gaussian scheme to integrate
over the MRN distribution. This keeps $npg$ to a reasonably low value
(typically $12$).%
}.

\subsection{Desorption processes}

Ejection can occur spontaneously (thermal evaporation) or by photo-desorption
or cosmic rays ejection. All processes are similar in the sense that
they only involve a single variable $N_{\mathrm{X}}(a)$ on a grain
of size $a$.

If the vibration frequency of the adsorbed particle is $\nu_{0}$,
the temperature of the grain is $T_{gr}(a)$ and the binding energy
is $T_{b}(\mathrm{X})$, then the number of evaporation per unit time
is:
\[
\nu_{0}\,\exp\left(-\frac{T_{b}(\mathrm{X})}{T_{g}(a)}\right)\, N_{\mathrm{X:}}(a)
\]

If the flux of photons (resp. cosmic rays) is $F_{ph}$ (resp. $F_{CR}$)
and the number of particles desorbed by impact is $\eta_{ph}$ (resp.
$\eta_{CR}$), then the number of desorption is (for a photon):
\[
F_{ph}\,\pi\, a^{2}\,\eta_{ph}\,\frac{d_{s}^{2}}{4\pi\, a^{2}}\, N_{\mathrm{X:}}(a)=F_{ph}\,\eta_{ph}\,\frac{d_{s}^{2}}{4}\, N_{\mathrm{X:}}(a)
\]

Writing $k_{ev}(a)=\nu_{0}\,\exp\left(-\frac{T_{b}(\mathrm{X})}{T_{g}(a)}\right)$,
$k_{ph}=F_{ph}\,\eta_{ph}\,\frac{d_{s}^{2}}{4}$ and $k_{CR}=F_{CR}\,\eta_{CR}\,\frac{d_{s}^{2}}{4}$,
we have:
\begin{equation}
-\frac{dN_{\mathrm{X:}}(a)}{dt}=\left(k_{ev}(a)+k_{ph}+k_{CR}\right)\, N_{\mathrm{X:}}(a)\label{eq_surf_ej}
\end{equation}
\begin{equation}
\frac{d[\mathrm{X}]}{dt}=\sum_{i}^{npg}\, k_{ev}(a_{i})\, w_{i}\, N_{\mathrm{X:}}(a_{i})+\left(k_{ph}+k_{CR}\right)\,\sum_{i}^{npg}\, w_{i}\, N_{\mathrm{X:}}(a_{i})\label{eq_eject}
\end{equation}

\subsection{Surface reactions}

We seek to compute the number of encounters per grain and per unit
time. Let us look at things from the point of view of $\mathrm{X\!\!:}$.
On a single grain of size $a$ the number of encounter per $\mathrm{s}$
is proportional to $1/t_{\mathrm{X}}$ the inverse hoping time of
$\mathrm{X\!\!:}$, the probability to find a $\mathrm{Y\!\!:}$ upon
landing and the number of $\mathrm{X\!\!:}$
\[
N_{\mathrm{X:}}(a)\,\frac{1}{t_{\mathrm{X}}}\,\frac{d_{s}^{2}}{4\pi\, a^{2}}\, N_{\mathrm{Y:}}(a)
\]

During the same time, from the point of view of $\mathrm{Y:}$, the
number of encounter made is
\[
N_{\mathrm{Y:}}(a)\,\frac{1}{t_{\mathrm{Y}}}\,\frac{d_{s}^{2}}{4\pi\, a^{2}}\, N_{\mathrm{X:}}(a)
\]

So the total number of encounters is:
\[
\frac{1}{2}\,\left(\frac{1}{t_{\mathrm{X}}}+\frac{1}{t_{\mathrm{Y}}}\right)\,\frac{d_{s}^{2}}{4\pi\, a^{2}}\, N_{\mathrm{X:}}(a)\, N_{\mathrm{Y:}}(a)
\]
where the factor of $\frac{1}{2}$ takes care of the fact that each
encounter has been counted twice. Thus, for two surface species, we
can write $k_{for}=\left(\frac{1}{t_{\mathrm{X}}(a)}+\frac{1}{t_{\mathrm{Y}}(a)}\right)\,\frac{d_{s}^{2}}{8\pi\, a^{2}}$,
and:
\begin{equation}
-\frac{dN_{\mathrm{X:}}(a)}{dt}=-\frac{dN_{\mathrm{Y:}}(a)}{dt}=k_{for}\, N_{\mathrm{X:}}(a)\, N_{\mathrm{Y:}}(a)\label{eq_surf_rxn}
\end{equation}

The production rate of $\mathrm{Z}$ in the gas phase, occurring directly
after the encounter of two adsorbed atoms, is obtained after integration
on the grain size distribution:
\begin{equation}
\frac{d[\mathrm{Z}]}{dt}=\frac{d_{s}^{2}}{8\pi}\,\sum_{i}^{npg}\left(\frac{1}{t_{\mathrm{X}}(a_{i})}+\frac{1}{t_{\mathrm{Y}}(a_{i})}\right)\,\frac{w_{i}}{a_{i}^{2}}\, N_{\mathrm{X:}}(a_{i})\, N_{\mathrm{Y:}}(a_{i})\label{eq_formLH}
\end{equation}

\subsection{Approximate $\mathrm{H}_{2}$ formation rate}

For a single grain size, and negligible photodesorption and cosmic
rays desorption, we can derive an analytic approximation to the $\mathrm{H}_{2}$
formation rate in the spirit of the discussion of \citet{biham:02}.
If $\mathrm{H}$ is the only atom sticking to a grain of size $a$
with a sticking probability of $1$, then:
\[
\frac{dN_{\mathrm{H:}}}{dt}=k_{ad}\,[\mathrm{H}]-\, k_{rej}\,[\mathrm{H}]\, N_{\mathrm{H:}}-k_{ev}\, N_{\mathrm{H:}}-2\, k_{for}\, N_{\mathrm{H:}}^{2}
\]

At steady state, this leads to:
\begin{equation}
N_{\mathrm{H:}}=\frac{[\mathrm{H}]}{[\mathrm{H}]_{b}}\,\left(1+\frac{[\mathrm{H}]_{a}}{[\mathrm{H}]}\right)\,\left[-1+\sqrt{1+2\, N_{\mathrm{H:}}^{max}\,\frac{[\mathrm{H}]_{b}}{[\mathrm{H}]}\,\frac{1}{\left(1+\frac{[\mathrm{H}]_{a}}{[\mathrm{H}]}\right)^{2}}}\right]\label{eq:LH_full}
\end{equation}
with $N_{\mathrm{H:}}^{max}=\frac{4\pi\, a^{2}}{d_{s}^{2}}$ the maximum
number of $\mathrm{H}$ on the grain, and the two critical densities
\textbf{$[\mathrm{H}]_{a}$} and\textbf{ }$[\mathrm{H}]_{b}$ defined
as:
\[
[\mathrm{H}]_{a}=\frac{k_{ev}}{k_{rej}}=\frac{4\,\nu_{0}}{\bar{v}\, d_{s}^{2}}\,\exp\left(-\frac{T_{b}}{T_{g}}\right)
\]

\[
[\mathrm{H}]_{b}=\frac{4\, k_{for}}{k_{rej}}=\frac{4\,\nu_{0}}{\overline{v}(\mathrm{H)}\, d_{s}^{2}}\,\frac{4}{N_{\mathrm{H:}}^{max}}\,\exp\left(-\frac{T_{d}}{T_{g}}\right)\,=\frac{4\,\nu_{0}}{\pi\,\overline{v}(\mathrm{H)\, a^{2}}}\,\mathrm{exp\left(-\frac{T_{d}}{T_{g}}\right)}
\]

Table~\ref{Table:appendix}gives the values of the critical densities
for different grain temperatures:

\begin{table*}
\caption{Critical densities in the LH formation rate of $\mathrm{H_{2}}$ for
amorphous carbon and a mean distance between physisorbed sites of
$2.6\,\textrm{\AA}$; $T$ is the gas temperature in $\mathrm{K}$.\label{Table:appendix}}
\centering

\begin{tabular}{lcccccc}
\hline 
\hline $T_{grain}\,(\mathrm{K})$ & \multicolumn{2}{c}{10} & \multicolumn{2}{c}{15} & \multicolumn{2}{c}{30}\tabularnewline
 & $a=10^{-6}\,\mathrm{cm}$ & $a=10^{-5}\,\mathrm{cm}$ & $a=10^{-6}\,\mathrm{cm}$ & $a=10^{-5}\,\mathrm{cm}$ & $a=10^{-6}\,\mathrm{cm}$ & $a=10^{-5}\,\mathrm{cm}$\tabularnewline
\hline 
$[\mathrm{H]_{a}}\,\mathrm{cm}^{-3}$ & $1.2\,10^{-5}/\sqrt{T}$ & $1.2\,10^{-5}/\sqrt{T}$ & $3.9\,10^{4}/\sqrt{T}$ & $3.9\,10^{4}/\sqrt{T}$ & $1.3\,10^{14}\,/\sqrt{T}$ & $1.3\,10^{14}/\sqrt{T}$\tabularnewline
$[\mathrm{H]_{b}}\,\mathrm{cm}^{-3}$ & $5.9\,10^{-3}/\sqrt{T}$ & $5.9\,10^{-1}/\sqrt{T}$ & $1.4\,10^{3}/\sqrt{T}$ & $1.4\,10^{5}/\sqrt{T}$ & $3.6\,10^{10}/\sqrt{T}$ & $3.6\,10^{12}/\sqrt{T}$\tabularnewline
\hline 
\end{tabular}
\end{table*}

The $\mathrm{H}_{2}$ formation rate per grain is then (with $n_{g}$
the number of grains per cubic centimeter of gas = $\frac{3\,\cdot1.4\,\cdot m_{\mathrm{H}}\,\cdot G}{4\pi\rho\, a^{3}}\, n_{\mathrm{H}}$):
\[
\frac{1}{n_{g}}\,\left.\frac{d[\mathrm{H}_{2}]}{dt}\right|_{LH}=k_{for}\, N_{\mathrm{H:}}^{2}
\]
 and can be given analytically from the previous formulae.\textbf{
$[\mathrm{H}]_{a}$} and\textbf{ }$[\mathrm{H}]_{b}$ vary slowly
with the gas temperature, but very strongly with the grain temperature.
Thus, we can define two limiting regimes for the gas phase atomic
hydrogen density:
\begin{itemize}
\item $[\mathrm{H}]\gg[\mathrm{H}]_{a}>[\mathrm{H}]_{b}$. Then:
\[
N_{\mathrm{H:}}=N_{\mathrm{H:}}^{max}
\]
\[
\frac{1}{n_{g}}\,\left.\frac{d[\mathrm{H}_{2}]}{dt}\right|_{LH}=\nu_{0}\,\frac{4\pi\, a^{2}}{d_{s}^{2}}\,\exp\left(-\frac{T_{d}}{T_{g}}\right)
\]

\end{itemize}
\[
\frac{d[\mathrm{H_{2}]}}{dt}\,=\,\nu_{0}\,\frac{1}{d_{s}^{2}}\,\times\,\frac{3\cdot1.4\cdot m_{\mathrm{H}}\cdot G}{\rho a}\,\times\, exp\left(-\frac{T_{d}}{T_{g}}\right)\,\times\, n_{\mathrm{H}}
\]
This is possible only for cold grains (typically below $15\,\mathrm{K}$).
So, it requires both a high density (or pressure) and low radiation
field.
\begin{itemize}
\item $[\mathrm{H}]\ll[\mathrm{H}]_{b}<[\mathrm{H}]_{a}$. Then:
\[
N_{\mathrm{H:}}=\frac{N_{\mathrm{H:}}^{max}}{[\mathrm{H}]_{a}}\,[\mathrm{H}]=\frac{\pi\, a^{2}\,\bar{v}}{\nu_{0}}\,\exp\left(\frac{T_{b}}{T_{g}}\right)\,[\mathrm{H}]
\]
\begin{equation}
\frac{1}{n_{g}}\,\left.\frac{d[\mathrm{H}_{2}]}{dt}\right|_{LH}=\pi\, a^{2}\,\bar{v}\,\frac{d_{s}^{2}\,\bar{v}}{4\,\nu_{0}}\,\exp\left(\frac{2\, T_{b}-T_{d}}{T_{g}}\right)\,[\mathrm{H}]^{2}\label{eq:LH_lowH}
\end{equation}

\end{itemize}
\[
\left.\frac{d[\mathrm{H}_{2}]}{dt}\right|_{LH}=\bar{v}(\mathrm{H)^{2}}\,\frac{d_{s}^{2}}{4\,\nu_{0}}\,\frac{3\times1.4\, m_{\mathrm{H}}\, G}{\pi\,\rho\, a}\,\exp\left(\frac{2\, T_{b}-T_{d}}{T_{g}}\right)\, n_{\mathrm{H}}\,[\mathrm{H}]^{2}
\]
This is the case for all grain sizes as soon as $T_{g}$ is above
about $25\,\mathrm{K}$. So it applies to all high radiation field
models. Here the formation rate grows as the square of the density
of $\mathrm{H}$.

The usual expression for the formation rate $R_{\mathrm{H_{2}}}$
in $cm^{3}s^{-1}$ follows from:
\[
\left.\frac{d[\mathrm{H}_{2}]}{dt}\right|_{LH}=R_{\mathrm{H}_{2}}\, n_{\mathrm{H}}\,[\mathrm{H}]
\]

Since $n_{g}$ is proportional to $n_{\mathrm{H}}$ we see that in
the first case $R_{\mathrm{H}_{2}}\propto1/[\mathrm{H}]$, whereas
in the second $R_{\mathrm{H}_{2}}\propto[\mathrm{H}]$. These relations
apply only with the approximations made here.

\section{Eley-Rideal mechanism\label{App_ER}}

\subsection{\textbf{Formalism}}

Let us consider the impact of a fast atom (hot gas) with a grain.
Since a full detailed description (taking into account all possible
kinds of surfaces) is much beyond the capacity of our model, we look
for an approximate mechanism that takes into account the following
constraints:
\begin{itemize}
\item It is efficient in ``hot'' gas and on ``hot'' grains. Therefore
the impinging $\mathrm{H}$ atom must eventually reach a chemisorbed
site on the grain.
\item It leads to $\mathrm{H}_{2}$ formation rates consistent with observational
constraints.
\item The number of free parameters is kept to the lowest possible number.
\end{itemize}
Since this process takes place at the edge of the cloud, we assume
that the grains are essentially bare (without ice coating) and that
the process does not depend on the grain temperature. This approximation
is justified since, at the edge of PDRs, grain temperatures (at most
$100\,\mathrm{K}$) are much lower than the gas temperature or chemical
binding energies on grain surfaces.

On impact, the gas phase $\mathrm{H}$ can find either a free chemisorpsion
site or an already chemisorbed $\mathrm{H}$. In the second case,
since the formation of $\mathrm{H}_{2}$ releases 4.5 eV, an energy
far higher than the chemisorbed well, we assume that a newly formed
$\mathrm{H}_{2}$ is immediately released in the gas phase. So:
\[
\left.\frac{d[\mathrm{H}_{2}]}{dt}\right|_{ER}=-\left.\frac{d[\mathrm{H::}]}{dt}\right|_{dest}=v_{th}\,<n\sigma_{gr}>\,[\mathrm{H}]\,\frac{[\mathrm{H::}]}{[\mathrm{H::}]_{max}}
\]
where $[\mathrm{H::}]_{max}$ is the maximum number of chemisorbed
$\mathrm{H}$ atoms (saturated grains) and $[\mathrm{H::}]$ the corresponding
abundance. $v_{th}$ is the thermal velocity of the gas phase $\mathrm{H}$
and we consider the geometrical cross-section to compute the total
amount of grain surface per unit volume $<n\sigma_{gr}>$. If the
mean distance between chemisorption sites is $d_{s}$ and is the same
on all types of grains, one can see (from purely geometric considerations)
that:
\[
\frac{<n\sigma_{gr}>}{[\mathrm{H::}]_{max}}=\frac{d_{s}^{2}}{4}
\]

This is true for any grain size distribution. If the gas phase atom
impacts a free chemisorption site, we have to estimate the probability
that it sticks to the grain. The simplest hypotheses requires that
it be proportional to the number of collisions of $\mathrm{H}$ with
grains per unit of time ($v_{th}\,<n\sigma_{gr}>$), possibly with
a barrier to cross $\exp\left(-\frac{T_{1}}{T}\right)$ (where $T$
is the gas temperature and $T_{1}$ the threshold), with a temperature
dependent sticking coefficient $\alpha(T)$ and proportional to the
``free room'' $\left(1-\frac{[\mathrm{H::}]}{[\mathrm{H::}]_{max}}\right)$.
So we write:
\[
\left.\frac{d[\mathrm{H::}]}{dt}\right|_{form}=\alpha(T)\, v_{th}\,<n\sigma_{gr}>\,[\mathrm{H}]\,\exp\left(-\frac{T_{1}}{T}\right)\,\left(1-\frac{[\mathrm{H::}]}{[\mathrm{H::}]_{max}}\right)
\]

This equation is split in the code in two: a direct formation reaction
and a ``pseudo'' rejection reaction. The corresponding rates are:
\[
\left.\frac{d[\mathrm{H::}]}{dt}\right|_{form,d}=\alpha(T)\, v_{th}\,<n\sigma_{gr}>\,\exp\left(-\frac{T_{1}}{T}\right)\,[\mathrm{H}]
\]
\[
\left.\frac{d[\mathrm{H::}]}{dt}\right|_{form,r}=-\alpha(T)\, v_{th}\,\frac{d_{s}^{2}}{4}\,\exp\left(-\frac{T_{1}}{T}\right)\,[\mathrm{H}]\,[\mathrm{H::}]
\]

\subsection{\textbf{Sticking coefficient and choice of $T_{1}$}}

There is not much information on how to define the sticking function
$\alpha(T)$, but we expect that it goes to $0$ for very high temperatures
(the atom just bounces on the grain without time to evacuate the excess
kinetic energy). We introduce the empirical form:
\begin{equation}
\alpha(T)=\frac{1}{1+\left(\frac{T}{T_{2}}\right)^{\beta}}\label{eq_stick_ER}
\end{equation}

In this expression, the index $\beta$ controls the steepness of the
decrease of $\alpha(T)$ and $T_{2}$ defines the temperature such
that $\alpha(T_{2})=\frac{1}{2}$.

We may constrain the value of $\beta$ by the following considerations:
\begin{itemize}
\item An estimate of the velocity $v_{2}$ above which the atom bounces
back to the gas is given by:
\[
v_{2}\sim d_{s}\,\nu_{0}
\]

\item The sticking coefficient is approximated as the fraction of gas phase
atom with velocity lower than $v_{2}$. Using a Maxwell distribution
at temperature $T$ we have:
\[
\alpha(T)\simeq\sqrt{\frac{2}{\pi}\,\left(\frac{m}{kT}\right)^{3}}\,\int_{0}^{v_{2}}v^{2}\,\exp\left(-\frac{m\, v^{2}}{2kT}\right)\, dv
\]
\[
=\mathrm{erf}\left(\sqrt{\frac{m\, v_{2}^{2}}{2kT}}\right)-\frac{2}{\sqrt{\pi}}\sqrt{\frac{m\, v_{2}^{2}}{2kT}}\,\exp\left(-\frac{m\, v_{2}^{2}}{2kT}\right)
\]

\end{itemize}
At low temperatures, this expression tends to $1$ as expected. At
high temperatures, we may expand this expression as a function of
$m\, v_{2}^{2}/2kT$. Defining $T_{2}$ by $v_{2}=\sqrt{\frac{8}{\pi}\,\frac{kT_{2}}{m}}$,
we get:
\[
\alpha(T)\simeq\frac{4}{3\sqrt{\pi}}\,\left(\frac{m}{2kT}\right)^{3/2}\, v_{2}^{3}=\frac{32}{3\pi^{2}}\,\left(\frac{T_{2}}{T}\right)^{3/2}
\]

This shows that $\beta=\frac{3}{2}$ in Eq~(\ref{eq_stick_ER}) is
exact. The prefactor is $1.08$, which we can take as $1$ given the
approximations involved; thus Eq~(\ref{eq_stick_ER}) is a good approximation
over the whole temperature range. The estimate of $v_{2}$ gives $T_{2}$
in the range $400-500\,\mathrm{K}$%
\footnote{The value $T_{2}=464\,\mathrm{K}$ was selected to match a previous
``guesstimate'' proportional to $10^{-4}\, T^{\beta}$. %
}. 

Given $\alpha(T)$, we may investigate which barrier $T_{1}$ gives
a ``standard'' formation rate of $3\,10^{-17}\,\mathrm{cm}^{3}\,\mathrm{s}^{-1}$
at a given temperature $T$. We find:
\[
T_{1}=T\,\log\left(\alpha(T)\,\left(\frac{2.8}{3}\,\sqrt{T}-1\right)\right)
\]

Postulating that where grains are warm, the gas is warm too, we may
require that this standard rate is reached for a temperature in the
range $[150:450]\,\mathrm{K}$. This translate into a range $[100:800]\,\mathrm{K}$
for $T_{1}$. Our choice of $T_{1}=300\,\mathrm{K}$ favors an efficient
formation, and reflects the idea that chemisorption is easy (but not
instantaneous) on grains with lots of surface defects. This leads
to a higher formation rate at high gas temperature as found observationally
by \citet{habart:04}.

\subsection{\textbf{Analytical approximation}}

\begin{figure}
\centering\includegraphics[width=1\columnwidth]{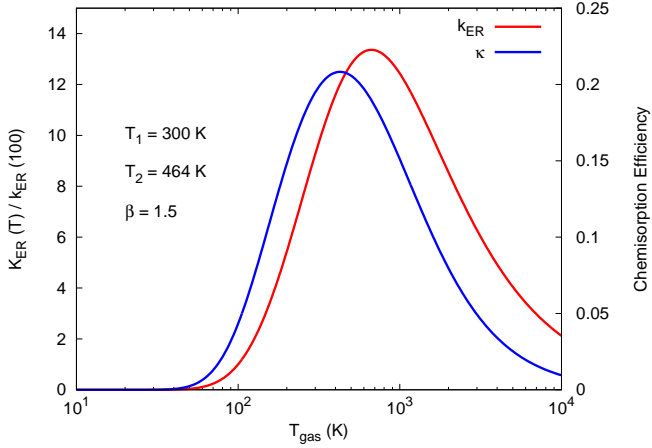}

\caption{Left axis: Variation of $k_{ER}$ with gas temperature $T$ (relative
to the one at $100\,\mathrm{K}$). Right axis: chemisorption efficiency
$\kappa$ (see text).\label{fig_ER_rate}}
\end{figure}

If hydrogen is the only chemisorbed species, the abundance of $\mathrm{H::}$
can be analytically derived at steady state:
\[
[\mathrm{H::}]=\frac{\alpha(T)\,\exp\left(-\frac{T_{1}}{T}\right)}{1+\alpha(T)\,\exp\left(-\frac{T_{1}}{T}\right)}\,[\mathrm{H::}]_{max}
\]

This leads to an $\mathrm{H}_{2}$ formation rate of:
\begin{equation}
\left.\frac{d[\mathrm{H}_{2}]}{dt}\right|_{ER}=v_{th}\,<n\sigma_{gr}>\,\kappa(T\text{)}\,[\mathrm{H}]=k_{ER}\,[\mathrm{H}]\, n_{\mathrm{H}}\label{eq:_ER_rate}
\end{equation}
with:
\[
\kappa(T\text{)}=\frac{\alpha(T)\,\exp\left(-\frac{T_{1}}{T}\right)}{1+\alpha(T)\,\exp\left(-\frac{T_{1}}{T}\right)}
\]

In this expression, $v_{th}\,<n\sigma_{gr}>\,[\mathrm{H}]$ refers
to a purely geometric collisional process. Fig.~\ref{fig_ER_rate}
displays the variation of the chemisorption rate as a function of
gas temperature (relative to the one at $100\,\mathrm{K}$). The rate
is negligible at low temperature (due to the exponential barrier).
It grows as the square root of $T$ once the barrier is negligible,
then is quenched by the sticking cutoff.\textbf{ }The chemisorption
efficiency, $\kappa(T\text{)}$, is displayed\textbf{ }on Fig.~\ref{fig_ER_rate}
(right axis). It peaks at a few hundred Kelvin and keeps significant
values up to a few thousands.

This behavior is qualitatively very similar to results found by \citet{cuppen:10}
(their Fig.~2) from Monte Carlo simulations of $\mathrm{H}_{2}$
formation including both physisorption and chemisorption. They find
also that formation efficiency increases for gas temperatures around
a thousand $\mathrm{K}$.

\begin{figure}
\centering\includegraphics[bb=65bp 65bp 365bp 275bp,clip,width=1\columnwidth]{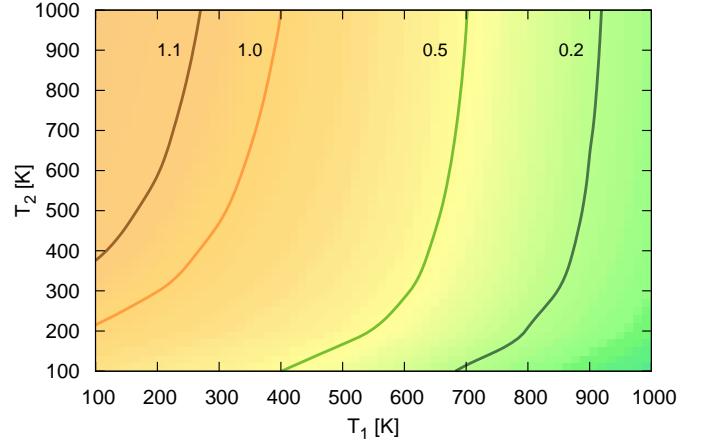}\caption{Effect of the variation of $T_{1}$ and $T_{2}$ on the intensity
of $\mathrm{H}_{2}$ 1-0 S(1) line for the model $P=10^{7}\,\mathrm{cm}^{-3}\,\mathrm{K}$
and $\chi=1000$. Values plotted in the plane $T_{1},T_{2}$ are $I(T_{1},T_{2})/I(300,464)$.\label{fig:T1T2var}}
\end{figure}

In Sect.~\ref{sub:Emission-lines} we present several line intensities
computed from models in which $\mathrm{H}_{2}$ is formed by ER and
LH mechanisms. In these models, we adopt $\beta=1.5$, $T_{1}=300\,\mathrm{K}$
and $T_{2}=464\,\mathrm{K}$. Fig.~\ref{fig:T1T2var} presents the
effect of variations of $T_{1}$ and $T_{2}$ on the intensity of
one line of $\mathrm{H}_{2}$. We note that $T_{1}$ is the most important
parameter and that the line intensity can be reduced by a factor $\simeq5$
if this parameter is increased from $300\,\mathrm{K}$ to $1000\,\mathrm{K}$.
For other lines, as 0-0 S(0) this decrease can reach a factor of $10$.
As mentioned above, for real interstellar grains, we can expect to
have a large range of $T_{1}$ depending on the nature and structure
of the grains surfaces. Even if most of chemisorbed sites have high
thresholds, it only requires a few low thresholds sites for the ER
mechanism to be efficient. 

\section{$A_{V}$ to size conversion\label{sec:Size_conv}}

For constant dust properties along the line of sight, it is possible
to convert optical depth in the visible to a distance (in $\mathrm{pc}$)
analytically. Let $C_{d}$ be the total proton column density to color
index ratio, and $R_{V}$ the usual extinction to color index ratio:
\[
C_{d}=\frac{N_{\mathrm{H}}}{E_{B-V}}\,;\quad R_{V}=\frac{A_{V}}{E_{B-V}}
\]

Then, using $A_{V}=2.5\,\log_{10}(e)\,\tau_{V}$, we have:
\[
N_{H}=\int_{0}^{l}n_{\mathrm{H}}\, ds=C_{d}\,\frac{E_{B-V}}{A_{V}}\,2.5\,\log_{10}(e)\,\int_{0}^{\tau_{V}^{max}}d\tau_{V}
\]
so, with $C_{t}=\frac{C_{d}}{R_{V}}\,2.5\,\log_{10}(e)$, we have:
\[
l=\int ds=C_{t}\,\int_{0}^{\tau_{V}^{max}}\frac{1}{n_{\mathrm{H}}}\, d\tau_{V}
\]

\bibliographystyle{aa}
\bibliography{H2_paper}

\begin{thebibliography}{66}
\expandafter\ifx\csname natexlab\endcsname\relax\def\natexlab#1{#1}\fi

\bibitem[{{Abrahamsson} {et~al.}(2007){Abrahamsson}, {Krems}, \&
  {Dalgarno}}]{abraham:07}
{Abrahamsson}, E., {Krems}, R.~V., \& {Dalgarno}, A. 2007, \apj, 654, 1171

\bibitem[{{Bachellerie} {et~al.}(2009){Bachellerie}, {Sizun}, {Aguillon},
  {Teillet-Billy}, {Rougeau}, \& {Sidis}}]{bachellerie:09}
{Bachellerie}, D., {Sizun}, M., {Aguillon}, F., {et~al.} 2009, Physical
  Chemistry Chemical Physics (Incorporating Faraday Transactions), 11, 2715

\bibitem[{{Bakes} \& {Tielens}(1994)}]{bakes:94}
{Bakes}, E.~L.~O. \& {Tielens}, A.~G.~G.~M. 1994, \apj, 427, 822

\bibitem[{{Biham} {et~al.}(2001){Biham}, {Furman}, {Pirronello}, \&
  {Vidali}}]{biham:01}
{Biham}, O., {Furman}, I., {Pirronello}, V., \& {Vidali}, G. 2001, \apj, 553,
  595

\bibitem[{{Biham} \& {Lipshtat}(2002)}]{biham:02}
{Biham}, O. \& {Lipshtat}, A. 2002, \pre, 66, 056103

\bibitem[{{Biham} {et~al.}(2005){Biham}, {Lipshtat}, \& {Perets}}]{biham:05}
{Biham}, O., {Lipshtat}, A., \& {Perets}, H.~B. 2005, in IAU Symposium, Vol.
  231, Astrochemistry: Recent Successes and Current Challenges, ed. D.~C.
  {Lis}, G.~A. {Blake}, \& E.~{Herbst}, 345--354

\bibitem[{{Black} \& {van Dishoeck}(1987)}]{black:87}
{Black}, J.~H. \& {van Dishoeck}, E.~F. 1987, \apj, 322, 412

\bibitem[{{Burgh} {et~al.}(2007){Burgh}, {France}, \& {McCandliss}}]{burgh:07}
{Burgh}, E.~B., {France}, K., \& {McCandliss}, S.~R. 2007, \apj, 658, 446

\bibitem[{{Burton} {et~al.}(2002){Burton}, {Londish}, \& {Brand}}]{burton:02}
{Burton}, M.~G., {Londish}, D., \& {Brand}, P.~W.~J.~L. 2002, \mnras, 333, 721

\bibitem[{{Cazaux} \& {Tielens}(2004)}]{cazaux:04}
{Cazaux}, S. \& {Tielens}, A.~G.~G.~M. 2004, \apj, 604, 222

\bibitem[{{Cazaux} \& {Tielens}(2010)}]{cazaux:10}
{Cazaux}, S. \& {Tielens}, A.~G.~G.~M. 2010, \apj, 715, 698

\bibitem[{{Chang} {et~al.}(2006){Chang}, {Cuppen}, \& {Herbst}}]{chang:06}
{Chang}, Q., {Cuppen}, H.~M., \& {Herbst}, E. 2006, \aap, 458, 497

\bibitem[{{Compi{\`e}gne} {et~al.}(2011){Compi{\`e}gne}, {Verstraete}, {Jones},
  {Bernard}, {Boulanger}, {Flagey}, {Le Bourlot}, {Paradis}, \&
  {Ysard}}]{compiegne:11}
{Compi{\`e}gne}, M., {Verstraete}, L., {Jones}, A., {et~al.} 2011, \aap, 525,
  A103+

\bibitem[{{Congiu} {et~al.}(2009){Congiu}, {Matar}, {Kristensen}, {Dulieu}, \&
  {Lemaire}}]{congiu:09}
{Congiu}, E., {Matar}, E., {Kristensen}, L.~E., {Dulieu}, F., \& {Lemaire},
  J.~L. 2009, \mnras, 397, L96

\bibitem[{{Cuppen} {et~al.}(2010){Cuppen}, {Kristensen}, \&
  {Gavardi}}]{cuppen:10}
{Cuppen}, H.~M., {Kristensen}, L.~E., \& {Gavardi}, E. 2010, \mnras, 406, L11

\bibitem[{{Cuppen} {et~al.}(2006){Cuppen}, {Morata}, \& {Herbst}}]{cuppen:06}
{Cuppen}, H.~M., {Morata}, O., \& {Herbst}, E. 2006, \mnras, 367, 1757

\bibitem[{{Draine} \& {Sutin}(1987)}]{draine:87}
{Draine}, B.~T. \& {Sutin}, B. 1987, \apj, 320, 803

\bibitem[{{Dubernet} {et~al.}(2006){Dubernet}, {Grosjean}, {Flower}, {Roueff},
  {Daniel}, {Moreau}, \& {Debray}}]{dubernet:06}
{Dubernet}, M., {Grosjean}, A., {Flower}, D., {et~al.} 2006, Journal of Plasma
  Research SERIES, Volume 7, p.~356-357, 7, 356

\bibitem[{{Duley} \& {Williams}(1993)}]{duley:93}
{Duley}, W.~W. \& {Williams}, D.~A. 1993, \mnras, 260, 37

\bibitem[{{Farebrother} {et~al.}(2000){Farebrother}, {Meijer}, {Clary}, \&
  {Fisher}}]{farebrother:00}
{Farebrother}, A.~J., {Meijer}, A.~J.~H.~M., {Clary}, D.~C., \& {Fisher}, A.~J.
  2000, Chemical Physics Letters, 319, 303

\bibitem[{{Fitzpatrick} \& {Massa}(1990)}]{fitzpatrick:90}
{Fitzpatrick}, E.~L. \& {Massa}, D. 1990, \apjs, 72, 163

\bibitem[{{Fitzpatrick} \& {Massa}(2007)}]{fitzpatrick:07}
{Fitzpatrick}, E.~L. \& {Massa}, D. 2007, \apj, 663, 320

\bibitem[{{Goicoechea} {et~al.}(2009){Goicoechea}, {Compi{\`e}gne}, \&
  {Habart}}]{Goicoechea09}
{Goicoechea}, J.~R., {Compi{\`e}gne}, M., \& {Habart}, E. 2009, \apjl, 699,
  L165

\bibitem[{{Gonzalez Garcia} {et~al.}(2008){Gonzalez Garcia}, {Le Bourlot}, {Le
  Petit}, \& {Roueff}}]{gonzalez:08}
{Gonzalez Garcia}, M., {Le Bourlot}, J., {Le Petit}, F., \& {Roueff}, E. 2008,
  \aap, 485, 127

\bibitem[{{Gry} {et~al.}(2002){Gry}, {Boulanger}, {Nehm{\'e}}, {Pineau des
  For{\^e}ts}, {Habart}, \& {Falgarone}}]{gry:02}
{Gry}, C., {Boulanger}, F., {Nehm{\'e}}, C., {et~al.} 2002, \aap, 391, 675

\bibitem[{{Habart} {et~al.}(2011){Habart}, {Abergel}, {Boulanger}, {Joblin},
  {Verstraete}, {Compi{\`e}gne}, {Pineau Des For{\^e}ts}, \& {Le
  Bourlot}}]{habart:11}
{Habart}, E., {Abergel}, A., {Boulanger}, F., {et~al.} 2011, \aap, 527, A122+

\bibitem[{{Habart} {et~al.}(2004){Habart}, {Boulanger}, {Verstraete},
  {Walmsley}, \& {Pineau des For{\^e}ts}}]{habart:04}
{Habart}, E., {Boulanger}, F., {Verstraete}, L., {Walmsley}, C.~M., \& {Pineau
  des For{\^e}ts}, G. 2004, \aap, 414, 531

\bibitem[{{Habart} {et~al.}(2005){Habart}, {Walmsley}, {Verstraete}, {Cazaux},
  {Maiolino}, {Cox}, {Boulanger}, \& {Pineau des For{\^e}ts}}]{habart:05}
{Habart}, E., {Walmsley}, M., {Verstraete}, L., {et~al.} 2005, \ssr, 119, 71

\bibitem[{{Hasegawa} \& {Herbst}(1993)}]{hasegawa:93}
{Hasegawa}, T.~I. \& {Herbst}, E. 1993, \mnras, 261, 83

\bibitem[{{Hasegawa} {et~al.}(1992){Hasegawa}, {Herbst}, \&
  {Leung}}]{hasegawa:92}
{Hasegawa}, T.~I., {Herbst}, E., \& {Leung}, C.~M. 1992, \apjs, 82, 167

\bibitem[{{Hauser} {et~al.}(1998){Hauser}, {Arendt}, {Kelsall}, {Dwek},
  {Odegard}, {Weiland}, {Freudenreich}, {Reach}, {Silverberg}, {Moseley},
  {Pei}, {Lubin}, {Mather}, {Shafer}, {Smoot}, {Weiss}, {Wilkinson}, \&
  {Wright}}]{hauser:98}
{Hauser}, M.~G., {Arendt}, R.~G., {Kelsall}, T., {et~al.} 1998, \apj, 508, 25

\bibitem[{{Hollenbach} {et~al.}(2009){Hollenbach}, {Kaufman}, {Bergin}, \&
  {Melnick}}]{hollenbach:09}
{Hollenbach}, D., {Kaufman}, M.~J., {Bergin}, E.~A., \& {Melnick}, G.~J. 2009,
  \apj, 690, 1497

\bibitem[{{Hollenbach} \& {Salpeter}(1971)}]{hollenbach:71}
{Hollenbach}, D. \& {Salpeter}, E.~E. 1971, \apj, 163, 155

\bibitem[{{Hollenbach} {et~al.}(1991){Hollenbach}, {Takahashi}, \&
  {Tielens}}]{hollenbach:91}
{Hollenbach}, D.~J., {Takahashi}, T., \& {Tielens}, A.~G.~G.~M. 1991, \apj,
  377, 192

\bibitem[{{Hornek{\ae}r} {et~al.}(2003){Hornek{\ae}r}, {Baurichter},
  {Petrunin}, {Field}, \& {Luntz}}]{hornekaer:03}
{Hornek{\ae}r}, L., {Baurichter}, A., {Petrunin}, V.~V., {Field}, D., \&
  {Luntz}, A.~C. 2003, Science, 302, 1943

\bibitem[{{Islam} {et~al.}(2010){Islam}, {Cecchi-Pestellini}, {Viti}, \&
  {Casu}}]{islam:10}
{Islam}, F., {Cecchi-Pestellini}, C., {Viti}, S., \& {Casu}, S. 2010, \apj,
  725, 1111

\bibitem[{{Islam} {et~al.}(2007){Islam}, {Latimer}, \& {Price}}]{islam:07}
{Islam}, F., {Latimer}, E.~R., \& {Price}, S.~D. 2007, \jcp, 127, 064701

\bibitem[{{Jenkins} \& {Shaya}(1979)}]{jenkins:79}
{Jenkins}, E.~B. \& {Shaya}, E.~J. 1979, \apj, 231, 55

\bibitem[{{Jenkins} \& {Tripp}(2001)}]{jenkins:01}
{Jenkins}, E.~B. \& {Tripp}, T.~M. 2001, \apjs, 137, 297

\bibitem[{{Jenkins} \& {Tripp}(2007)}]{jenkins:07}
{Jenkins}, E.~B. \& {Tripp}, T.~M. 2007, in Astronomical Society of the Pacific
  Conference Series, Vol. 365, SINS - Small Ionized and Neutral Structures in
  the Diffuse Interstellar Medium, ed. {M.~Haverkorn \& W.~M.~Goss}, 51--+

\bibitem[{{Jura}(1974)}]{jura:74}
{Jura}, M. 1974, \apj, 191, 375

\bibitem[{{Katz} {et~al.}(1999){Katz}, {Furman}, {Biham}, {Pirronello}, \&
  {Vidali}}]{katz:99}
{Katz}, N., {Furman}, I., {Biham}, O., {Pirronello}, V., \& {Vidali}, G. 1999,
  \apj, 522, 305

\bibitem[{{Kaufman} {et~al.}(2006){Kaufman}, {Wolfire}, \&
  {Hollenbach}}]{kaufmann:06}
{Kaufman}, M.~J., {Wolfire}, M.~G., \& {Hollenbach}, D.~J. 2006, \apj, 644, 283

\bibitem[{{Kim} {et~al.}(2011){Kim}, {Balgar}, \& {Hasselbrink}}]{kim:11}
{Kim}, H., {Balgar}, T., \& {Hasselbrink}, E. 2011, Chemical Physics Letters,
  508, 1

\bibitem[{{Latimer} {et~al.}(2008){Latimer}, {Islam}, \& {Price}}]{latimer:08}
{Latimer}, E.~R., {Islam}, F., \& {Price}, S.~D. 2008, Chemical Physics
  Letters, 455, 174

\bibitem[{{Launay} \& {Roueff}(1977)}]{launay:77}
{Launay}, J.~M. \& {Roueff}, E. 1977, \aap, 56, 289

\bibitem[{{Le Bourlot} {et~al.}(1995{\natexlab{a}}){Le Bourlot}, {Pineau des
  Forets}, {Roueff}, {Dalgarno}, \& {Gredel}}]{lebourlot:95b}
{Le Bourlot}, J., {Pineau des Forets}, G., {Roueff}, E., {Dalgarno}, A., \&
  {Gredel}, R. 1995{\natexlab{a}}, \apj, 449, 178

\bibitem[{{Le Bourlot} {et~al.}(1995{\natexlab{b}}){Le Bourlot}, {Pineau des
  Forets}, {Roueff}, \& {Flower}}]{lebourlot:95}
{Le Bourlot}, J., {Pineau des Forets}, G., {Roueff}, E., \& {Flower}, D.~R.
  1995{\natexlab{b}}, \aap, 302, 870

\bibitem[{{Le Petit} {et~al.}(2009){Le Petit}, {Barzel}, {Biham}, {Roueff}, \&
  {Le Bourlot}}]{lepetit:09}
{Le Petit}, F., {Barzel}, B., {Biham}, O., {Roueff}, E., \& {Le Bourlot}, J.
  2009, \aap, 505, 1153

\bibitem[{{Le Petit} {et~al.}(2006){Le Petit}, {Nehm{\'e}}, {Le Bourlot}, \&
  {Roueff}}]{lepetit:06}
{Le Petit}, F., {Nehm{\'e}}, C., {Le Bourlot}, J., \& {Roueff}, E. 2006, \apjs,
  164, 506

\bibitem[{{Lemaire} {et~al.}(2010){Lemaire}, {Vidali}, {Baouche}, {Chehrouri},
  {Chaabouni}, \& {Mokrane}}]{lemaire:10}
{Lemaire}, J.~L., {Vidali}, G., {Baouche}, S., {et~al.} 2010, \apjl, 725, L156

\bibitem[{{Lipshtat} \& {Biham}(2003)}]{lipshtat:03}
{Lipshtat}, A. \& {Biham}, O. 2003, \aap, 400, 585

\bibitem[{{Lipshtat} {et~al.}(2004){Lipshtat}, {Biham}, \&
  {Herbst}}]{lipshtat:04}
{Lipshtat}, A., {Biham}, O., \& {Herbst}, E. 2004, \mnras, 348, 1055

\bibitem[{{Lohmar} \& {Krug}(2006)}]{lohmar06}
{Lohmar}, I. \& {Krug}, J. 2006, \mnras, 370, 1025

\bibitem[{{Lohmar} {et~al.}(2009){Lohmar}, {Krug}, \& {Biham}}]{lohmar09}
{Lohmar}, I., {Krug}, J., \& {Biham}, O. 2009, \aap, 504, L5

\bibitem[{{Mathis} {et~al.}(1983){Mathis}, {Mezger}, \& {Panagia}}]{mathis:83}
{Mathis}, J.~S., {Mezger}, P.~G., \& {Panagia}, N. 1983, \aap, 128, 212

\bibitem[{{Mathis} {et~al.}(1977){Mathis}, {Rumpl}, \& {Nordsieck}}]{mathis:77}
{Mathis}, J.~S., {Rumpl}, W., \& {Nordsieck}, K.~H. 1977, \apj, 217, 425

\bibitem[{{Rachford} {et~al.}(2009){Rachford}, {Snow}, {Destree}, {Ross},
  {Ferlet}, {Friedman}, {Gry}, {Jenkins}, {Morton}, {Savage}, {Shull},
  {Sonnentrucker}, {Tumlinson}, {Vidal-Madjar}, {Welty}, \&
  {York}}]{rachford09}
{Rachford}, B.~L., {Snow}, T.~P., {Destree}, J.~D., {et~al.} 2009, \apjs, 180,
  125

\bibitem[{{Sakong} \& {Kratzer}(2010)}]{sakong:10}
{Sakong}, S. \& {Kratzer}, P. 2010, \jcp, 133, 054505

\bibitem[{{Sch{\"o}ier} {et~al.}(2005){Sch{\"o}ier}, {van der Tak}, {van
  Dishoeck}, \& {Black}}]{schoier:05}
{Sch{\"o}ier}, F.~L., {van der Tak}, F.~F.~S., {van Dishoeck}, E.~F., \&
  {Black}, J.~H. 2005, \aap, 432, 369

\bibitem[{{Sheffer} {et~al.}(2008){Sheffer}, {Rogers}, {Federman}, {Abel},
  {Gredel}, {Lambert}, \& {Shaw}}]{Sheffer:08}
{Sheffer}, Y., {Rogers}, M., {Federman}, S.~R., {et~al.} 2008, \apj, 687, 1075

\bibitem[{{Sheffer} {et~al.}(2011){Sheffer}, {Wolfire}, {Hollenbach},
  {Kaufman}, \& {Cordier}}]{sheffer:11}
{Sheffer}, Y., {Wolfire}, M.~G., {Hollenbach}, D.~J., {Kaufman}, M.~J., \&
  {Cordier}, M. 2011, \apj, 741, 45

\bibitem[{{Sizun} {et~al.}(2010){Sizun}, {Bachellerie}, {Aguillon}, \&
  {Sidis}}]{sizun:10}
{Sizun}, M., {Bachellerie}, D., {Aguillon}, F., \& {Sidis}, V. 2010, Chemical
  Physics Letters, 498, 32

\bibitem[{{van Dishoeck}(1988)}]{vd:88}
{van Dishoeck}, E.~F. 1988, {Photodissociation and photoionization processes},
  49--72

\bibitem[{{Weingartner} \& {Draine}(2001)}]{weingartner:01}
{Weingartner}, J.~C. \& {Draine}, B.~T. 2001, \apjs, 134, 263

\bibitem[{{Wolfire} {et~al.}(1995){Wolfire}, {Hollenbach}, {McKee}, {Tielens},
  \& {Bakes}}]{wolfire:95}
{Wolfire}, M.~G., {Hollenbach}, D., {McKee}, C.~F., {Tielens}, A.~G.~G.~M., \&
  {Bakes}, E.~L.~O. 1995, \apj, 443, 152

\end{thebibliography}

\end{document}